\documentstyle[psfig]{l-aa}

\begin{document}

\thesaurus{3(11.06.2; 11.16.1; 11.19.2; 11.19.6; 11.19.7)}

\title{The {\it z}-structure of disk galaxies towards the galaxy
planes\thanks{Based on observations obtained at the European Southern
Observatory, La Silla, Chile}}

\author{R. de Grijs, R.F. Peletier\thanks{Current address: Durham
University, Physics Dept., South Road, Durham, DH1 3LE, United Kingdom}
\and P.C. van der Kruit}

\institute{Kapteyn Astronomical Institute, University of Groningen,
P.O. Box 800, 9700 AV Groningen, The Netherlands}

\offprints{R. de Grijs; grijs@astro.rug.nl}

\date{Received date; accepted date}

\maketitle

\markboth{R. de Grijs et al.: The {\it z}-structure of disk galaxies}{}

\begin{abstract}
We present a detailed study of a statistically complete sample of highly
inclined disk galaxies in the near-infrared $K'$ band.  Since the
$K'$-band light is relatively insensitive to contamination by galactic
dust, we have been able to follow the vertical light distributions all
the way down to the galaxy planes.  \\ The mean levels for the sharpness
of the $K'$-band luminosity peaks indicate that the vertical luminosity
distributions are more peaked than expected for the intermediate
sech({\it z}) distribution, but rounder than exponential.  After fitting
a generalized family of fitting functions characterised by an exponent
2/n ($n = \infty$ for exponential, $n = 2$ for sech and $n = 1$ for
sech$^2$; van der Kruit 1988) we find that the mean value for 2/{\it n}
in the $K'$ band equals $<$2/{\it n}$>_{K'} = 0.538, \sigma_{K'} =
0.198$.  Since projection of not completely edge-on galaxies onto the
plane of the sky causes vertical luminosity profiles to become rounder,
we have performed simulations that show that it is possible that all our
galaxies can have intrinsically exponential vertical surface brightness
distributions.  \\ We find that the profile shape is independent of
galaxy type, and varies little with position along the major axis.  The
fact that we observe this in all our sample galaxies indicates that the
formation process of the galaxy disks perpendicular to the galaxy planes
is a process intrinsic to the disks themselves. 
\end{abstract}

\section{Introduction}

\subsection{Motivation}

It is well-known from optical observations that many disk galaxies show
a universal, exponential behaviour of the vertical luminosity away from
the galaxy planes.  This observation, and the assumption that the
mass-to-light ratio in the disk is constant, leads us to believe that
galaxies have a universal, exponential vertical mass density profile
(e.g., Tsikoudi 1979; van der Kruit \& Searle 1981a). 

This nearly universal vertical structure is likely the result of
secular, internal evolution (van der Kruit \& Searle 1981a; Carlberg
1987).  As Carlberg (1987) states, the continuous variation of stellar
kinematics from the youngest to the oldest disk stars strongly suggests
that an ongoing dynamical evolution is indeed present. 

We know from observations of our Galaxy that galaxy disks probably
consist of multiple components of increasing age, velocity dispersion
and scale height; the density is the sum of all these components. 
An exponentially decreasing density distribution with distance from the
galaxy plane therefore puts interesting constraints on the star
formation rate (SFR) and the dynamical evolution of the disk.  Burkert
\& Yoshii (1996) have shown, that these exponential vertical density
profiles are a natural result of disk evolution if gaseous protodisks
settle into isothermal equilibrium prior to star formation.  If the star
formation and cooling rates are comparable, stellar exponential {\it
z}-profiles arise due to the gravitational contraction of the gas
towards the galaxy plane. 

\subsection{Observational status}

Several models have been proposed and tested to account for the vertical
distributions observed close to the galaxy planes.

Van der Kruit \& Searle (1981a,b, 1982a,b) studied the surface
brightness distributions of edge-on disk galaxies in optical passbands,
which were significantly affected by dust contamination.  From these
studies, they proposed the self-gravitating isothermal sheet (Spitzer
1942) as a model for the description of the vertical light distribution:

\begin{equation}
L(z) = L_0 \hbox{ sech}^2 (z/z_0)
\end{equation}
where $L_0$ is the surface luminosity in the plane of the galaxy, $z_0$
is the vertical scale parameter and {\it z} is the distance from the
galaxy plane, respectively.  At large {\it z} heights, the vertical
scale heights are equal for both the isothermal model and the
exponential approximation, where $z_0$ equals twice the exponential
scale height.  At intermediate {\it z}, most galaxies are dominated by
the thin disk luminosity, which is thought to be locally isothermal (van
der Kruit \& Searle 1981a). 

The relations among the values for the density in the plane, $\rho(0)$,
the vertical scale parameter, $z_0$, and the vertical velocity
dispersion $\langle v_z^2 \rangle^{1/2}$ for the solar neighbourhood
show that the disk of our Galaxy is self-gravitating (van der Kruit \&
Searle 1981a).  Based on the quite similar values for $z_0$ found by
Van der Kruit \& Searle (1981a) they argued that it may be reasonable to
assume that the disks of other galaxies are also self-gravitating. 

Near-infrared observations of edge-on disk galaxies have shown an excess
of light over the isothermal model at small distances from the galaxy
planes, where the optical photometry is strongly affected by dust
absorption (e.g., Wainscoat et al. 1989; Aoki et al. 1991; van Dokkum
et al. 1994). Wainscoat et al. (1989) show that the {\it
z}-dependence of the light in the large southern edge-on IC 2531
demonstrates a more strongly peaked profile than expected from the
isothermal sheet model, which appears to be better fitted by an
exponential:

\begin{equation}
L(z) = L_0 \exp (-z/h_z) ,
\end{equation}
(where $h_z$ is the exponential vertical scale height), although their
limited resolution does not unambiguously differentiate between the models. 
For the vertical {\it K}-band light distribution in NGC 891, Aoki et al. 
(1991) find that the exponential model fits the data remarkably well up
to those {\it z}-distances where the seeing convolution becomes
significant. 

\subsection{An intermediate solution}

Although the exponential model is mathematically attractive because of
its simplicity, there is no firm physical basis for such a model.  An
exponential vertical mass density distribution can be constructed by
adding up multiple stellar disk components.  This can only be done if
the contributions from stars with larger velocity dispersions are
increasingly dominating with increasing distance from the galaxy plane. 
However, a mechanism to account for such a process is as yet unknown
(Burkert \& Yoshii 1996). 

As van der Kruit (1988) argues, a pure exponential distribution also has
some undesired properties, the most important one being a sharp minimum
of the velocity dispersion in the plane.  Fuchs \& Wielen's (1987)
results show moderate gradients, much smaller than required for the
exponential distribution (Bahcall 1984a,b).  Therefore, van der Kruit
(1988) proposed that an intermediate distribution, such as the
``sech({\it z})'' distribution, could be a more appropriate one to use:

\begin{equation}
L(z) = L_0 \hbox{ sech} (z/h_z) ,
\end{equation}
to account for the deviations from an isothermal sheet in the galaxy
planes. 

Wainscoat et al.'s (1989) near-infrared photometry of IC 2531 would
agree with this model, as would star counts for our Galaxy (e.g.,
Gilmore \& Reid 1983). 

\subsection{The Galaxy}

Studies of the Galaxy provide valuable information on the vertical
structure of galaxy disks.  These studies, based on star counts, have
the advantage over the studies of external galaxies that they are less
affected by dust absorption (in the solar neighbourhood) and effects of
the presence of a young stellar population.  Moreover, studies of the
vertical luminosity structure in our Galaxy benefit greatly from the
higher spatial resolution compared to that in external galaxies. 
Gilmore \& Reid (1983) and Pritchet (1983) conclude that the stellar
{\it z}-distribution in our Galaxy is better approximated by an
exponential rather than an isothermal profile. 

On the other hand, Hill et al. (1979) derived density laws for A and F
dwarfs towards the North Galactic Pole, which cannot be fit well by an
exponential distribution, although they may be approximated as such in
short distance bins. Although they find that the F stars are roughly
consistent with a single exponential, the A stars can only be
approximated by a single exponential closer to the Galactic Plane and
with a significantly smaller scale height than the F dwarfs. 

Based on observations in the near-infrared, Kent et al. (1991) concluded
that the vertical light distribution (and hence probably also the mass
distribution) follows an exponential law more closely than an
isothermal sheet approximation.  However, they did not compare the
observed light distribution to other, intermediate models.  

\subsection{Near-Infrared Observations}

The study of edge-on galaxies in the near-infrared is valuable to reveal
the true stellar distributions, as the near-infrared wavelengths permit
to study these even at small {\it z}. 

In this paper we study the vertical luminosity profiles of a
statistically complete sample of edge-on disk galaxies in the $K'$ band. 
This sample is the largest of its kind available at the moment.  For
that reason we are able to study the structure of edge-on galaxies in a
statistically consistent way, even down to very small distances from the
galaxy planes.  

The $K'$-band light (which is comparable to the standard Johnson {\it K}
band) is likely to be dominated by the old disk, and the young disk
contribution is relatively unimportant. Moreover, the mass-to-light ratio
is almost independent of metallicity and age in this band.  Rix \& Rieke
(1993) find, from monitoring the gravity-sensitive CO(2.3 $\mu$m) index
in the large disk galaxy M51, that young red supergiants do not distort
the {\it K}-band image significantly.  They find that at most small
portions of the spiral arms at {\it K} have large contributions from
young stars. 

The {\it K}-band wavelength is too short for a substantial amount of
direct emission from the dust.  The high dust temperatures required to
emit in {\it K} (800 -- 1000 $^\circ$K) are associated only with young
stellar objects and compact H{\sc ii} regions (see, e.g., Wainscoat et
al. 1989).

The light in {\it K} is dominated by giants, which constitute only a
small fraction of the stellar mass. However, old population giant
stars have the same spatial distribution as the main-sequence stars
(Rix \& Rieke 1993).

Therefore, since neither dust nor young, luminous red stars strongly
affect the {\it K}-band image, {\it K}-band imaging with infrared arrays
is a reliable and efficient method to map surface mass variations
through surface brightness variations (Rix \& Rieke 1993)

In Sect. \ref{approach.sec} we present the sample properties and
describe the data reduction method used. The results of our detailed
analysis of the vertical profiles are presented in Sect.
\ref{results.sec}, in which we also compare the results obtained in the
optical {\it I} band to those from the $K'$ band observations. In Sect.
\ref{discussion.sec} we discuss the main observational results in the
context of star formation and global galaxy structure parameters.
Finally, we summarize and conclude the paper in Sect. \ref{summary.sec}.

\section{Approach}
\label{approach.sec}

\subsection{Sample selection, observations and data reduction}

To study the structural parameters of edge-on spiral galaxies we
selected a statistically complete sample taken from the Surface
Photometry Catalogue of the ESO-Uppsala Galaxies (ESO-LV; Lauberts \&
Valentijn 1989) with the following properties:
\begin{itemize}
\item their inclinations are greater than or equal to 87$^\circ$;
\item the angular blue diameters ($D_{25}$) are larger than $2.'2$;
\item the galaxy types range from S0 to Sd, and
\item they should be non-interacting.
\end{itemize}

The inclinations were determined following Guthrie (1992), assuming a
true axial ratio $\log R_0 = 0.95$, corresponding to an intrinsic
flattening $q_0 = (b/a)_0$ of 0.11.  From this intrinsic flattening the
inclinations {\it i} were derived by using Hubble's (1926) formula

\begin{equation}
\cos^2 i = (q^2 - q_0^2)/(1 - q_0^2),
\end{equation}
where $q = b/a$ is the observed axis ratio.

Of the total sample of 93 southern edge-ons, an arbitrary subsample of
24 galaxies was observed in the near-infrared $K'$ band in two observing
runs of 4 and 3 nights, respectively.  The selection of these 24
observed sample galaxies depended solely on the allocation of telescope
time; the galaxies cover the southern sky rather uniformly.  

By applying a $V/V_{\rm max}$ completeness test (e.g., Davies 1990;
de Jong \& van der Kruit 1994) we derived that the ESO-LV is
statistically complete for diameter-limited samples with $D_{25}^B \ge
1.'0$.  To check the completeness of our subsample, we calculated, based
on a limiting diameter $D_{25}^B \ge 2.'2$, that $V/V_{\rm max} =
0.502 \pm 0.253$, which implies statistical completeness. 

The near-infrared observations were obtained with the IRAC2B camera at
the ESO/MPI 2.2m telescope of the European Southern Observatory (ESO) in
Chile.  The IRAC2B camera is equipped with a Rockwell 256$\times$256
pixel NICMOS3 HgCdTe array.  For both observing runs, in July 1994 and
January 1995, we used the IRAC2B camera with Objective C, corresponding
to a pixel size of $0.''491$ (40 $\mu$m) and a field of view of
$125''\times125''$. 

At both runs we used the $K'$ filter available at ESO (central
wavelength $\lambda_{\rm c} = 2.15 \mu$m, bandpass $\Delta \lambda =
0.32 \mu$m).  We chose to observe in $K'$ rather than in {\it K} band
(with $\lambda_{\rm c} = 2.2 \mu$m, and $\Delta \lambda = 0.40
\mu$m), since the $K'$ band is almost as little affected by dust as the
{\it K} band, but has a lower sky background (Wainscoat \& Cowie 1992). 

We took sky images and object frames alternately, both with equal
integration times (in sequences of 12 $\times$ 10s), and spatially
separated by $\sim 5'$. 

Supplementary observations in the Thuan \& Gunn (1976) {\it I} band were
obtained during a number of observing runs at ESO.  Most of the {\it
I}-band observations were obtained with the Danish 1.54m telescope,
equipped with a 1081$\times$1040 pixel TEK CCD with a pixel size of
24$\mu$m (0.36 $''$/pix).  The field of view thus obtained is
$6'.5\times6.'2$.  The TEK CCD was used in slow read-out mode in order
to decrease the pixel-to-pixel noise.  The Thuan \& Gunn (1976) {\it
I}-band characteristics match those of a Johnson {\it I} filter (Buser
1978).  For all observed galaxies we determined the colour terms
required for the calibration to the Cousins system using standard stars. 

Gaps in the observed sample were filled in by service observations with
the Dutch 0.92m telescope, equipped with a 512$\times$512 pixel TEK CCD. 
It has a pixel size of 27 $\mu$m (0.44 $''$/pix), corresponding to a
field of view of $3.'9\times3.'9$. 

Both telescopes were used in direct imaging mode, at prime focus. 
Details of the specific observations can be found in Table
\ref{chap7obs.tab}. 

{
\begin{table}
\caption[ ]{\label{chap7obs.tab}{\bf Log of the {\it I} and $K'$-band
observations}
\newline Columns: (1) Galaxy name (ESO-LV); (2) Telescope
used (Dan 1.5 = Danish 1.54m; Dut 0.9 = Dutch 0.92m; ESO 2.2 = ESO/MPI
2.2m); (3) Date of observation (ddmmyy) (4) Passband observed in; (5)
Exposure time in seconds; (in the $K'$ band the same integration time was
spent to observe sky images) (6) Seeing FWHM in arcsec}

\begin{flushleft}
\begin{tabular}{llclrc}
\hline
\hline
Galaxy & Telescope & Date & Band   & Exp.time & Seeing \\
~~~(1) & ~~~~(2)   & (3)  & ~~(4) & (5)~~~~ & (6)    \\
\hline 
026G-06 & ~Dan 1.5 & 090794 & ~~$I$  & 2$\times$900          & 1.6 \\
        & ~ESO 2.2 & 110794 & ~~$K'$ & 2$\times$12$\times$10 & 1.5 \\
033G-22 & ~Dan 1.5 & 120194 & ~~$I$  & 2$\times$900          & 0.9 \\
041G-09 & ~Dan 1.5 & 100794 & ~~$I$  & 2$\times$900          & 1.4 \\
        & ~ESO 2.2 & 090794 & ~~$K'$ & 3$\times$12$\times$10 & 1.5 \\
074G-15 & ~Dan 1.5 & 110794 & ~~$I$  & 2$\times$900          & 1.3 \\
138G-14 & ~Dan 1.5 & 080794 & ~~$I$  & 2$\times$900          & 1.6 \\
141G-27 & ~Dan 1.5 & 090794 & ~~$I$  & 2$\times$900          & 1.2 \\
        & ~ESO 2.2 & 090794 & ~~$K'$ & 2$\times$12$\times$10 & 1.4 \\
142G-24 & ~Dan 1.5 & 110794 & ~~$I$  & 2$\times$900          & 1.4 \\
        & ~ESO 2.2 & 090794 & ~~$K'$ & 2$\times$12$\times$10 & 1.4 \\
157G-18 & ~Dan 1.5 & 100194 & ~~$I$  & 2$\times$900          & 1.2 \\
        & ~ESO 2.2 & 240295 & ~~$K'$ & 2$\times$12$\times$10 & 1.1 \\
201G-22 & ~Dan 1.5 & 090194 & ~~$I$  & 2$\times$900          & 1.2 \\
        & ~ESO 2.2 & 240295 & ~~$K'$ & 2$\times$12$\times$10 & 1.1 \\
202G-35 & ~Dut 0.9 & 061094 & ~~$I$  & 2$\times$900          & 1.2 \\
235G-53 & ~Dan 1.5 & 120794 & ~~$I$  &  600                  & 1.5 \\
        & ~Dut 0.9 & 170396 & ~~$I$  & 2$\times$1200         & 1.3 \\
240G-11 & ~Dut 0.9 & 051094 & ~~$I$  & 2$\times$1200         & 1.6 \\
        &          & 061094 & ~~$I$  & 2$\times$1200         & 1.6 \\
263G-15 & ~Dan 1.5 & 090194 & ~~$I$  &  900                  & 1.0 \\
        &          & 100194 & ~~$I$  &  600                  & 1.1 \\
        &          & 120194 & ~~$I$  &  900                  & 1.0 \\
        & ~ESO 2.2 & 240295 & ~~$K'$ &          12$\times$10 & 1.4 \\
        &          & 250295 & ~~$K'$ &          12$\times$10 & 1.4 \\
263G-18 & ~Dut 0.9 & 190194 & ~~$I$  & 2$\times$900          & 1.2 \\
269G-15 & ~Dut 0.9 & 130396 & ~~$I$  & 2$\times$1200         & 1.3 \\
280G-13 & ~Dut 0.9 & 140396 & ~~$I$  & 2$\times$1200         & 1.4 \\
286G-18 & ~Dan 1.5 & 090794 & ~~$I$  & 2$\times$900          & 1.3 \\
        & ~ESO 2.2 & 090794 & ~~$K'$ &          12$\times$10 & 1.6 \\
        &          & 110794 & ~~$K'$ &          12$\times$10 & 1.2 \\
288G-25 & ~Dut 0.9 & 051094 & ~~$I$  & 2$\times$1200         & 1.3 \\
311G-12 & ~Dan 1.5 & 120194 & ~~$I$  & 2$\times$900          & 1.0 \\
        & ~ESO 2.2 & 240295 & ~~$K'$ & 2$\times$12$\times$10 & 0.9 \\
315G-20 & ~Dan 1.5 & 090194 & ~~$I$  & 2$\times$900          & 1.7 \\
        & ~ESO 2.2 & 240295 & ~~$K'$ & 2$\times$12$\times$10 & 1.1 \\
321G-10 & ~Dut 0.9 & 280493 & ~~$I$  &  300                  & 1.2 \\
322G-73 & ~Dut 0.9 & 230396 & ~~$I$  & 2$\times$1200         & 1.2 \\
322G-87 & ~Dut 0.9 & 270493 & ~~$I$  &  300                  & 1.3 \\
340G-08 & ~Dan 1.5 & 100794 & ~~$I$  & 2$\times$900          & 1.3 \\
340G-09 & ~Dut 0.9 & 140396 & ~~$I$  & 2$\times$1200         & 1.4 \\
        & ~ESO 2.2 & 090794 & ~~$K'$ & 2$\times$12$\times$10 & 1.5 \\
358G-26 & ~Dan 1.5 & 120194 & ~~$I$  & 2$\times$900          & 1.0 \\
358G-29 & ~Dan 1.5 & 100194 & ~~$I$  & 2$\times$900          & 1.3 \\
        & ~ESO 2.2 & 260295 & ~~$K'$ & 2$\times$12$\times$10 & 1.5 \\
377G-07 & ~Dut 0.9 & 100396 & ~~$I$  & 2$\times$1200         & 1.2 \\
383G-05 & ~Dan 1.5 & 080794 & ~~$I$  & 2$\times$900          & 1.3 \\
        & ~ESO 2.2 & 090794 & ~~$K'$ & 2$\times$12$\times$10 & 1.1 \\
416G-25 & ~Dan 1.5 & 130194 & ~~$I$  & 2$\times$900          & 1.0 \\
        & ~ESO 2.2 & 260295 & ~~$K'$ & 2$\times$12$\times$10 & 1.4 \\
435G-14 & ~Dut 0.9 & 201293 & ~~$I$  & 2$\times$900          & 1.0 \\
        & ~ESO 2.2 & 240295 & ~~$K'$ & 2$\times$12$\times$10 & 1.5 \\
\hline
\end{tabular}
\end{flushleft}
\end{table}
}

\addtocounter{table}{-1}
{
\begin{table}
\caption[ ]{(Continued)}

\begin{flushleft}
\begin{tabular}{llclrc}
\hline
\hline
Galaxy & Telescope & Date & Band   & Exp.time & Seeing \\
~~~(1) & ~~~~(2)   & (3)  & ~~(4) & (5)~~~~ & (6)    \\
\hline 
435G-25 & ~Dut 0.9 & 230493 & ~~$I$  &  600                  & 1.7 \\
        &          & 231293 & ~~$I$  & 2$\times$900          & 1.3 \\
        &          & 060194 & ~~$I$  & 2$\times$900          & 1.2 \\
        & ~ESO 2.2 & 240295 & ~~$K'$ & 2$\times$12$\times$10 & 1.6 \\
435G-50 & ~Dan 1.5 & 100194 & ~~$I$  & 2$\times$900          & 1.1 \\
437G-62 & ~Dut 0.9 & 040195 & ~~$I$  & 2$\times$1200         & 1.0 \\
        & ~ESO 2.2 & 260295 & ~~$K'$ & 2$\times$12$\times$10 & 1.1 \\
444G-21 & ~Dut 0.9 & 230396 & ~~$I$  & 2$\times$1200         & 1.2 \\
446G-18 & ~Dan 1.5 & 090794 & ~~$I$  & 2$\times$900          & 1.1 \\
        & ~ESO 2.2 & 090794 & ~~$K'$ & 2$\times$12$\times$10 & 1.3 \\
        &          & 120794 & ~~$K'$ & 2$\times$12$\times$10 & 1.0 \\
446G-44 & ~Dut 0.9 & 140396 & ~~$I$  & 2$\times$1200         & 1.2 \\
        & ~ESO 2.2 & 090794 & ~~$K'$ & 2$\times$12$\times$10 & 1.4 \\
460G-31 & ~Dan 1.5 & 080794 & ~~$I$  & 2$\times$900          & 1.5 \\
        & ~ESO 2.2 & 090794 & ~~$K'$ & 2$\times$12$\times$10 & 1.5 \\
487G-02 & ~Dan 1.5 & 090194 & ~~$I$  & 2$\times$900          & 1.1 \\
        & ~ESO 2.2 & 240295 & ~~$K'$ & 2$\times$12$\times$10 & 1.1 \\
500G-24 & ~Dan 1.5 & 130194 & ~~$I$  & 2$\times$900          & 1.0 \\
        & ~ESO 2.2 & 250295 & ~~$K'$ &  2$\times$4$\times$30 & 1.6 \\
505G-03 & ~Dut 0.9 & 010496 & ~~$I$  & 2$\times$1200         & 1.3 \\
506G-02 & ~Dut 0.9 & 280493 & ~~$I$  &  300                  & 1.0 \\
509G-19 & ~Dan 1.5 & 100794 & ~~$I$  & 2$\times$900          & 1.0 \\
        & ~ESO 2.2 & 100794 & ~~$K'$ & 2$\times$12$\times$10 & 1.2 \\
531G-22 & ~Dut 0.9 & 270994 & ~~$I$  & 2$\times$1200         & 1.5 \\
555G-36 & ~Dan 1.5 & 130194 & ~~$I$  & 2$\times$900          & 1.1 \\
564G-27 & ~Dut 0.9 & 211293 & ~~$I$  & 2$\times$900          & 1.4 \\
        &          & 221293 & ~~$I$  & 2$\times$900          & 1.3 \\
        & ~ESO 2.2 & 240295 & ~~$K'$ & 2$\times$12$\times$10 & 0.9 \\
575G-61 & ~Dut 0.9 & 230396 & ~~$I$  & 2$\times$1200         & 1.2 \\
\hline
\end{tabular}
\end{flushleft}
\end{table}
}

During the reduction of the near-infrared observations, each sky frame
was compared with the two sky frames taken nearest in time in order to
detect stars in the sky frames.  These stars were filtered out by using
a median filter and thus the resulting cleaned sky images are very
similar to the actual sky contributions. 

To circumvent the effects of bad pixels and to obtain accurate
flatfielding we moved the object across the array between subsequent
exposures. Therefore, for most galaxies mosaicing of either 4 or 8
image frames was required to obtain complete galaxy images. The
mosaicing was done by using common stars in the frames to determine the
exact spatial offsets. In the rare case that no common stars could be
determined, we used the telescope offsets as our mosaicing offsets. The
overlapping area was used to determine the adjustment of sky levels
needed by means of a least squares fit. 

Bad pixels and bad areas on the array were masked out and not considered
during the entire reduction process.  Only after mosaicing was finished,
the areas that still did not contain any valid observations were
interpolated by a 2-dimensional linear plane fit (see Peletier [1993]
for a detailed description of the reduction method used). 

The calibration of the near-IR observations was done by using the
SAAO/ESO/ISO Faint Standard Stars (Carter \& Meadows 1995).  We used
the corrections published by Wainscoat \& Cowie (1992) to transform the
$K'$ measurements to the {\it K} band.  The accuracy of the $K'$-band
zero-point offsets we could reach was $\sim 0.08$ mag at both observing
runs. The limiting factors here were flatfielding errors.

The {\it I}-band images were reduced following standard reduction
procedures (see de Grijs \& van der Kruit 1996); for the calibration of
these observations Landolt fields were used (Landolt 1992).  The {\it
I}-band calibration could be done to an accuracy of $\sim 0.03 - 0.05$
mag, depending on the telescope and observing run.

Both our {\it I}-band observations and the $K'$-band data were taken at
photometric (parts of) nights.

\subsection{Vertical profiles}

We extracted vertical luminosity profiles at a number of positions along
the major axes of the sample galaxies.  A semi-logarithmic binning
algorithm was applied to the galaxies both radially and vertically, in
order to retain an approximately constant overall signal-to-noise (S/N)
ratio in the resulting vertical profiles.  Since the most significant
differences between our models become clear at small {\it z}, no
vertical binning was applied close to the galaxy planes.

We rejected those profiles with low S/N ratios (generally the outermost
profiles) and those that were clearly affected by artifacts in the data
or foreground stars.  For all galaxies we have been able to sample the
vertical light distribution at various positions along the major axis
outside the region where the bulge contribution dominates.  For galaxies
of types $T \ge -1.0$ we could determine these distributions for at
least 4 of these independent positions. 

The positions of the galaxy planes were determined by folding the
vertical profiles and under the assumption of symmetrical light
distributions with respect to the planes, in the near-infrared $K'$-band
observations. 

In Fig. \ref{colmaps.fig} we present the {\it I--K} colour maps of the
galaxies discussed in this paper.

\begin{figure*}
\caption[]{\label{colmaps.fig}Calibrated {\it I--K} colour maps of our
sample galaxies, displayed on the same scale.  Each colour map has been
inserted in a rectangular frame of size $80''\times200''$.  Galaxy
images were rotated from their original sky orientations so that the
major axes lie vertically.  The corresponding calibrated {\it I--K}
colours are indicated by the gray-scale bar at the bottom of the figure. 
{\it (a)} ESO026G-06; {\it (b)} ESO041G-09; {\it (c)} ESO141G-27; {\it
(d)} ESO142G-24; {\it (e)} ESO157G-18; {\it (f)} ESO201G-22; {\it (g)}
ESO263G-15; {\it (h)} ESO286G-18; {\it (i)} ESO311G-12; {\it (j)}
ESO315G-20; {\it (k)} ESO340G-09; {\it (l)} ESO358G-29; {\it (m)}
ESO383G-05; {\it (n)} ESO416G-25; {\it (o)} ESO435G-14; {\it (p)}
ESO435G-25; {\it (q)} ESO437G-62; {\it (r)} ESO446G-18; {\it (s)}
ESO446G-44; {\it (t)} ESO460G-31; {\it (u)} ESO487G-02; {\it (v)}
ESO500G-24; {\it (w)} ESO509G-19; {\it (x)} ESO564G-27.}
\end{figure*}

\subsection{Comparison to published luminosity profiles}

Surface brightness profiles of edge-on galaxies observed with modern
detectors are scarcely available in the recent literature. 

\begin{figure}
\psfig{figure=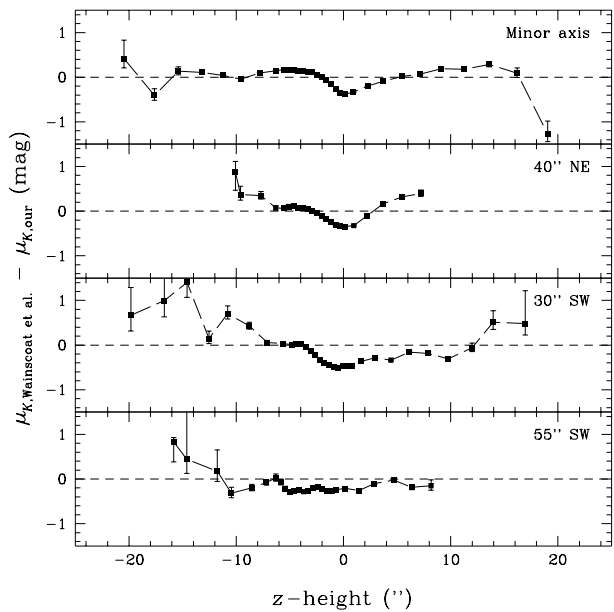,width=8.5cm}
\caption[]{\label{wainscoat.fig}Comparison of {\it K}-band vertical
profiles obtained by Wainscoat et al. (1989) and our observations. To
obtain these profiles we averaged strips of $7.''5$ wide.}
\end{figure}

In the near-infrared {\it K} band we compared our observations of the
large southern edge-on galaxy ESO 435G-25 with those of Wainscoat et al. 
(1989), obtained using a raster scan technique with an aperture of
5$''$.  Although our observations of ESO 435G-25 are of a much higher
quality and were taken with a much higher resolution, we find remarkably
good agreement between Wainscoat et al.'s (1989) and our {\it K}-band
observations, as can be seen in Fig.  \ref{wainscoat.fig}.  We have
extracted vertical profiles from our calibrated {\it K}-band image at
exactly the same positions and in the same way as was done by Wainscoat
et al.  (1989).  The distinct drop in the difference profiles at the
position of the galaxy plane is caused by the difference in resolution
between both data sets. 

A large and homogeneous set of {\it I}-band observations of southern
Sb--Sc galaxies has recently been published by Mathewson et al.  (1992)
and Mathewson \& Ford (1996).

\begin{figure*}
\psfig{figure=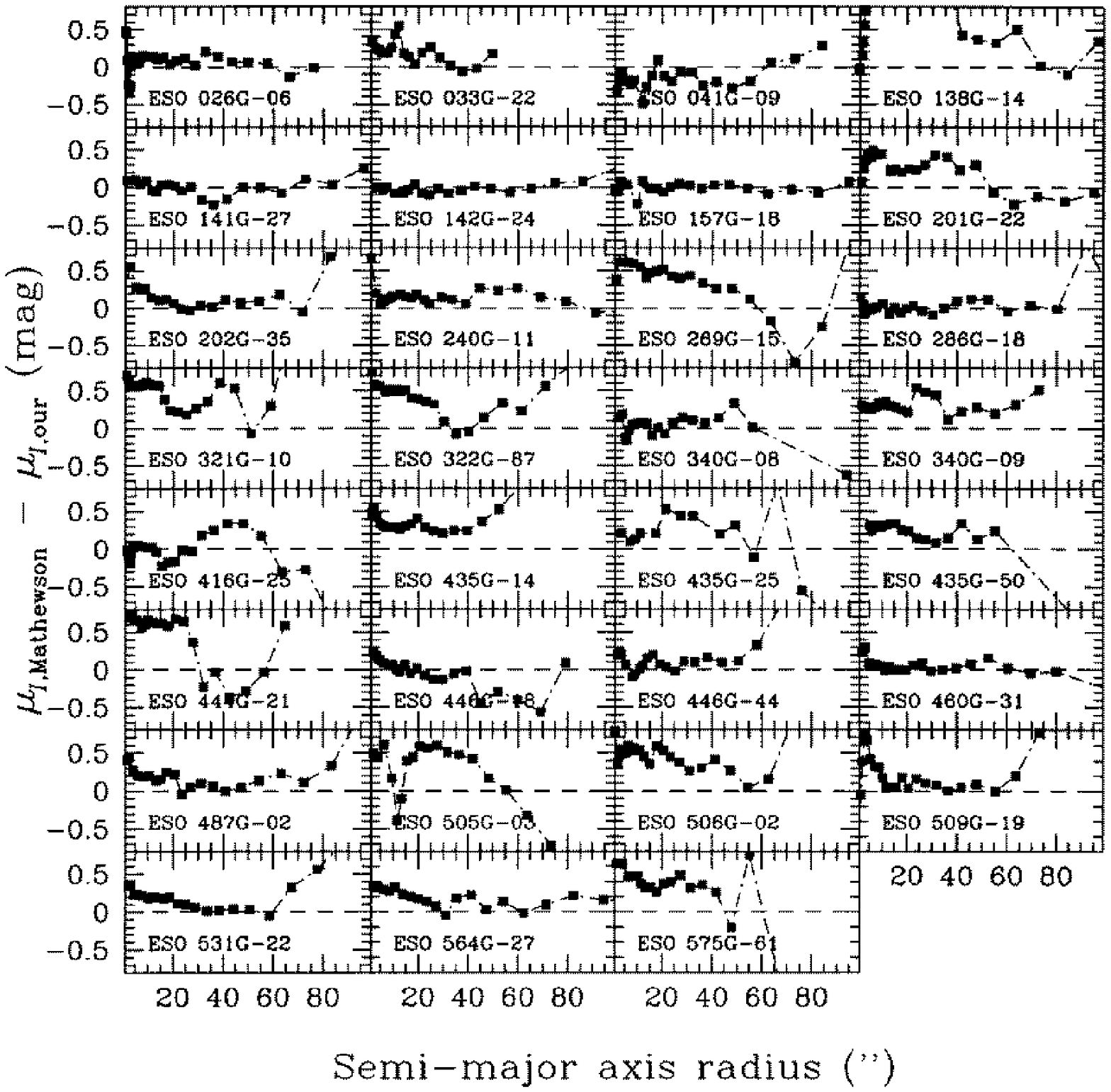,width=19cm}
\caption[]{\label{mathewson.fig}Comparison between azimuthally averaged
{\it I}-band profiles published by Mathewson et al. (1992) and Mathewson
\& Ford (1996), and those obtained from fitting ellipses to our
observations.}
\end{figure*}

Azimuthally averaged luminosity profiles were obtained by fitting
ellipses to the galaxy isophotes, whose intensity, ellipticity and
position angle were allowed to vary with each ellipse.  Bad pixels,
cosmic rays and foreground stars were masked out, so that these would
not affect the results from the ellipse-fitting routine.  Although this
method works sufficiently well for low and moderately-inclined galaxies,
when dealing with highly-inclined or edge-on galaxies the ellipse
fitting is severely influenced by the presence of a central dust lane
and the non-elliptical outer galaxy isophotes.  Unfortunately, since
Mathewson et al.  (1922) and Mathewson \& Ford (1996) did not tabulate
the ellipticities nor the position angles used for the individual
ellipses obtained for each galaxy, we can at best compare azimuthally
averaged profiles which were obtained with the same free parameters.  A
comparison between the azimuthally averaged {\it I}-band luminosity
profiles of Mathewson et al.  (1992) and Mathewson \& Ford (1996) and
those obtained from our observations is shown in Fig.  \ref{mathewson.fig}. 

In general, we find that the differences between our and Mathewson's
measurements are small, although clear deviations are appreciated in a
number of cases.  In particular for those galaxies for which the
difference between our and Mathewson's profiles is relatively large
(e.g., ESO 138G-14), we used observations obtained on different nights
or with a different telescope to check our results.  It was found that
the features shown in Fig.  \ref{mathewson.fig} can be reproduced to
within the observational errors.  The main cause of deviations between
Mathewson's and our profiles, in particular at small semi-major axis
radii, is the unpredictable influence of dust, which greatly affects the
ellipse fitting in the inner galaxy regions. 

From this comparison to previously published surface brightness
profiles, we conclude that our observations reproduce both the
photometric zero points and the behaviour of the galaxy light as a
function of position across the galaxy disk to within the observational
errors. 

\subsection{Extinction correction}

As was shown by Aoki et al. (1991), the inner contours of the {\it
K}-band image of NGC 891 are asymmetric. This shows that absorption is
not completely negligible, even at these near-infrared wavelengths.

To correct for this small absorption effect at {\it K}, we used the
{\it I--K} colour index as an extinction indicator, following Knapen et al. 
(1995). We assumed that deviations from an average {\it I--K} colour are
due to dust only and that the Galactic extinction law applies also in
external galaxies (see, e.g., Jansen et al. 1994). Although the exact
use of this law requires a detailed knowledge of the geometry of the
mixture of dust and stars, any errors caused by our assumptions are of
second order, since the {\it K} extinction due to dust is small. 

As an example, in NGC 891, Aoki et al. (1991) pointed out that the
southwest part of the {\it K}-band image is rather patchy along the
major axis, compared to the northeast part. The most likely explanation
for this patchiness is that it is due to dust associated with spiral
arms, as can be deduced from colour images.

In our assumptions, we ignore two kinds of systematic errors, as was
pointed out by Knapen et al.  (1995): a contribution of smoothly
distributed dust, which does not alter the galaxy's morphology, and
effects due to population changes, which are believed to be small.  In
general I-K colours of stellar populations are very similar.  For
example stellar population models (e.g., Vazdekis et al. 1996) only
show a small range for models of different ages and metallicities. 
Although population gradients probably are present, the errors made by
not taking them into account are likely to be so small that it is better
to apply this extinction correction than not to apply it. 

Summarized, we corrected our profiles as follows:

\begin{equation}
K_{\rm corr} = K_{\rm obs} - {1 \over A_I / A_K} \Bigl( (I-K)_{\rm obs}
- (I-K)_{\rm avg}) \Bigr) ,
\end{equation}
where $K_{\rm corr}$ and $K_{\rm obs}$ are the corrected and observed
profiles, $(I-K)_{\rm obs}$ and $(I-K)_{\rm avg}$ are the observed and
mean $(I-K)$ colours, and $A_I$ and $A_K$ are the extinction in {\it I}
and {\it K}, respectively.  We used $A_I / A_K = 4.30$ (Rieke \&
Lebofsky 1985).  This extinction correction amounts to 0.55 mag in {\it
K} at maximum, thus showing that we are dealing with optically thin
regions in our galaxies.  Therefore, by applying this correction, the
errors that are caused by the incorrect underlying assumption that the
dust is distributed in a foreground screen, are small. 

\subsection{A generalized family of density laws}

By analyzing these $K'$-band images of our sample of edge-on disk
galaxies we should be able to distinguish statistically between the
various models for the vertical luminosity and mass density
distribution.

As it seems reasonable to take the isothermal and exponential
distributions as the two extremes, van der Kruit (1988) proposed to use
the family of density laws

\begin{equation}
\label{family.eq}
K(z) = 2^{-2/n} K_0 \hbox{ sech}^{2/n} (nz/2z_0) , \qquad (n > 0)
\end{equation}
where {\it K(z)} is the observed {\it K}-band vertical density profile,
$K_0$ is the extrapolated outer surface density in the galaxy plane,
{\it z} is the distance from the plane, and $z_0$ is the
vertical scale parameter. The isothermal model is the extreme
for {\it n} = 1, and the exponential is the other extreme for $n =
\infty$. For comparison, for the isothermal sheet $z_0 = 2 h_z$, where
$h_z$ is the exponential vertical scale height. Therefore, the sharpness
of the peak in a luminosity profile is determined by the exponent 2/{\it
n} of this family of density laws (\ref{family.eq}).

\begin{figure}
\psfig{figure=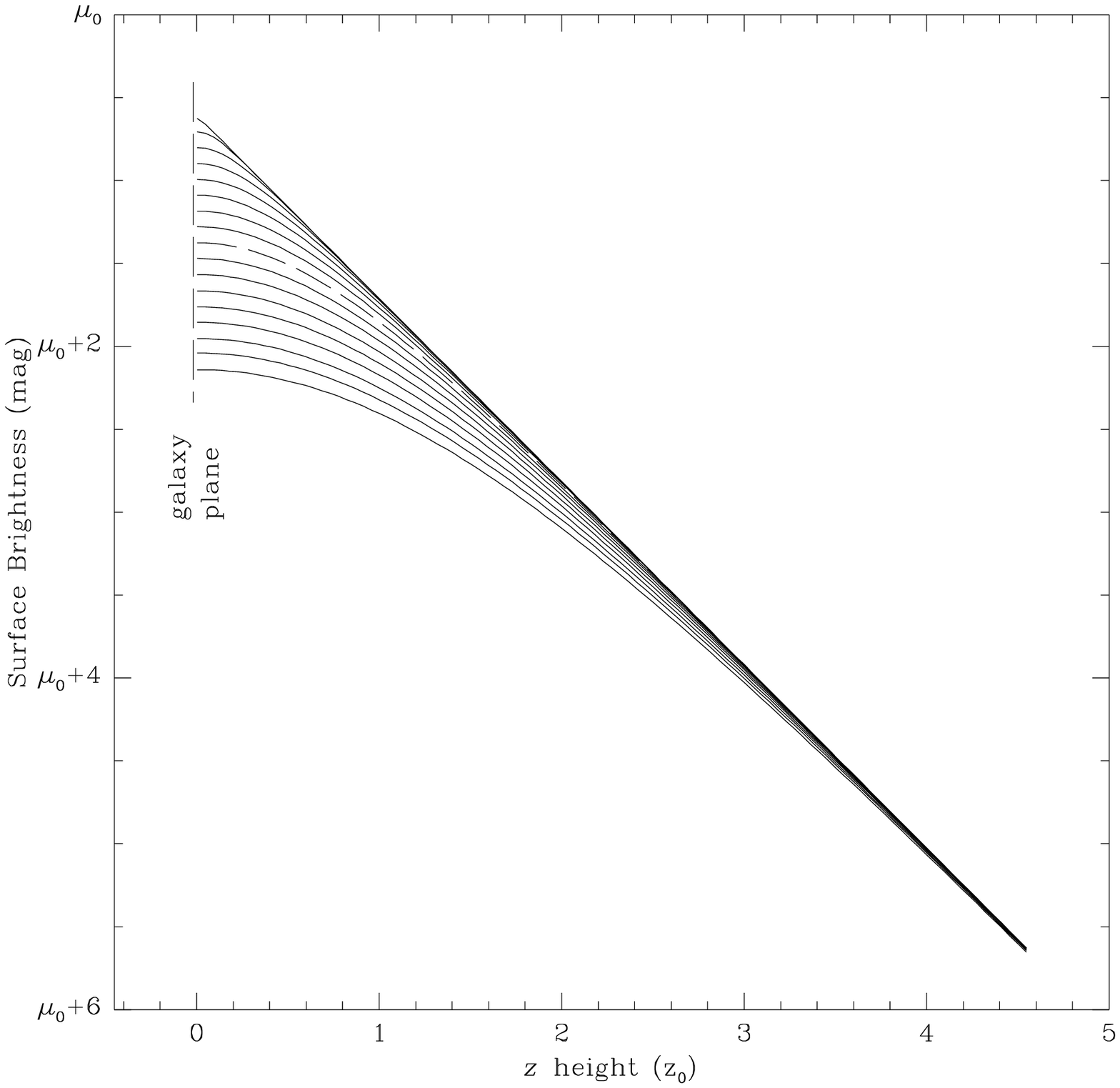,width=9cm}
\caption{}{\label{models.fig}The family of density (luminosity) laws
(\ref{family.eq}) with the isothermal (2/{\it n} = 2.0) and the
exponential (2/{\it n} = 0.0) distributions as the two extremes.  The
difference in 2/{\it n} between two successive model distributions is
0.125; for clarity, the sech({\it z}) model distribution (2/{\it n} =
1.0) is shown as the dashed profile ($\mu_0$ is the central surface
brightness).}
\end{figure}

In Fig.  \ref{models.fig} we plot a few model luminosity distributions,
with the exponential and the isothermal functions as the two extremes. 
We have adopted identical central surface brightnesses and vertical
scale heights for each of these models. 

\section{Results}
\label{results.sec}

\subsection{Near-infrared profiles}

In order to quantitatively distinguish between the various models, we
tried to fit the extinction-corrected vertical light profiles in the
{\it K} band with the extrapolated central surface brightness, the
vertical scale parameter {\em and} the exponent 2/{\it n} as free
parameters.  To check the validity of the best-fitting 2/{\it n} thus
obtained, we also tried to fit the vertical luminosity profiles with a
continuous series of fixed values for the 2/{\it n} parameter.  We
compared the minimum in the $\chi^2$ distribution as a function of
2/{\it n} with the best-fitting {\it n} value resulting from the
three-parameter fits.  From this comparison, we concluded that resulting
2/{\it n} values from the three-parameter fits could reproduce the
minima in the $\chi^2$ distribution with sufficient accuracy to be
reliable indicators of the profiles' shapes. 

To show the accuracy of our fits, in Fig.  \ref{fitprofs.fig} we have
compiled all our vertical {\it K}-band profiles.  We show the
best-fitting model luminosity distributions obtained with the
three-parameter fitting routine for those profiles that are sufficiently
symmetric.  The vertical luminosity profiles for each galaxy are shown
in two separate panels, representing either side of the galaxy with
respect to the galaxy center; for comparison, the central profiles are
shown in each panel.  In Table \ref{positions.tab} 
we have listed the positions along the major axis at which the vertical
luminosity profiles were extracted. 

\begin{figure*} 
\hspace*{-1cm}
\psfig{figure=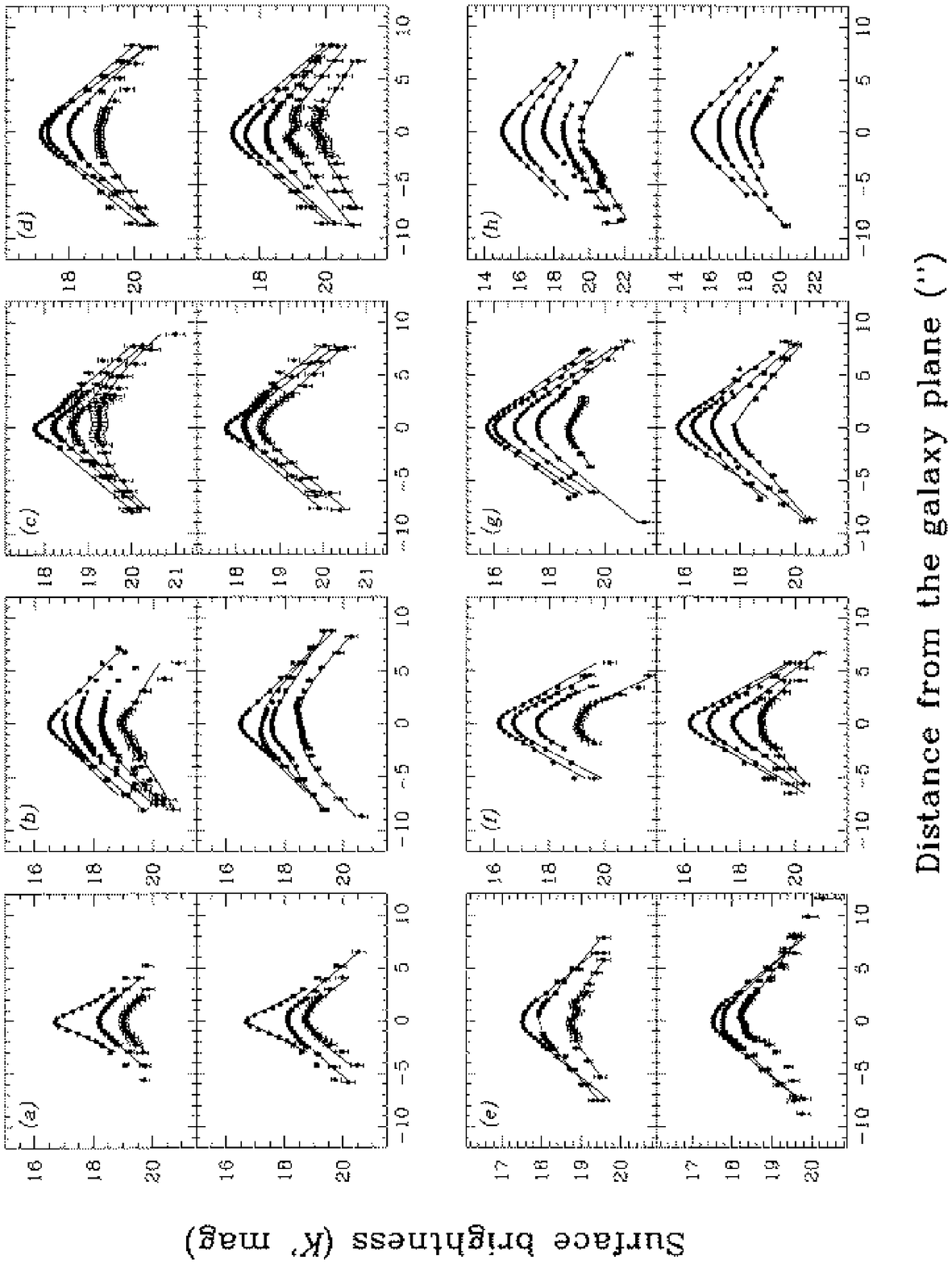}
\caption{}{\label{fitprofs.fig}Vertical {\it K}-band luminosity profiles
for the sample galaxies: {\it (a)} ESO026G-06; {\it (b)} ESO041G-09;
{\it (c)} ESO141G-27; \\ 
{\it (d)} ESO142G-24; {\it (e)} ESO157G-18; {\it (f)} ESO201G-22; {\it
(g)} ESO263G-15; {\it (h)} ESO286G-18.} 
\end{figure*}

\begin{figure*}
\hspace*{-1cm}
\addtocounter{figure}{-1}
\psfig{figure=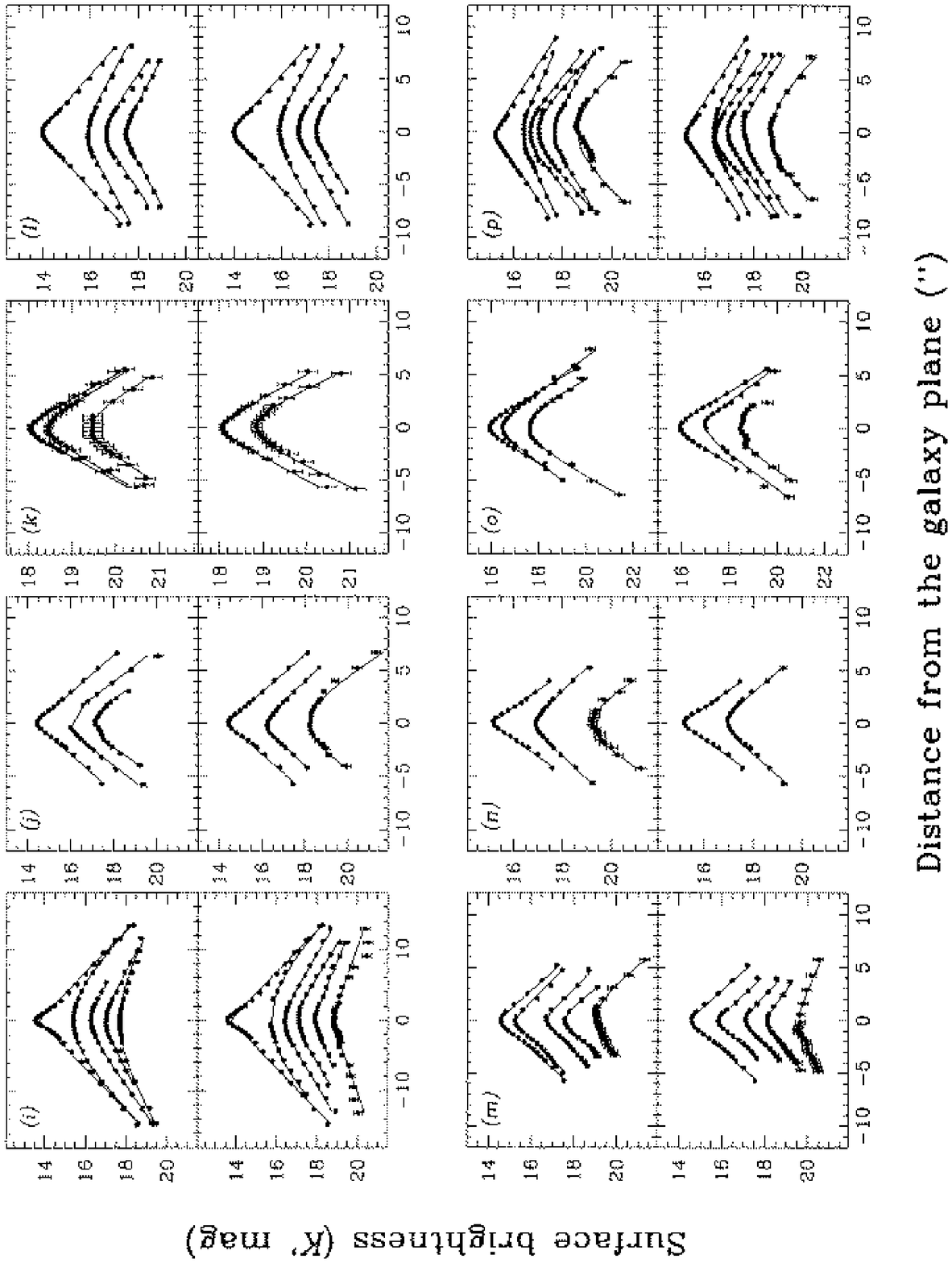}
\caption{}{(Continued) {\it (i)} ESO311G-12; {\it (j)} ESO315G-20; {\it
(k)} ESO340G-09; {\it (l)} ESO358G-29; {\it (m)} ESO383G-05; {\it (n)}
ESO416G-25; \\ {\it (o)} ESO435G-14; {\it (p)} ESO435G-25.} 
\end{figure*}

\begin{figure*}
\hspace*{-1cm}
\addtocounter{figure}{-1}
\psfig{figure=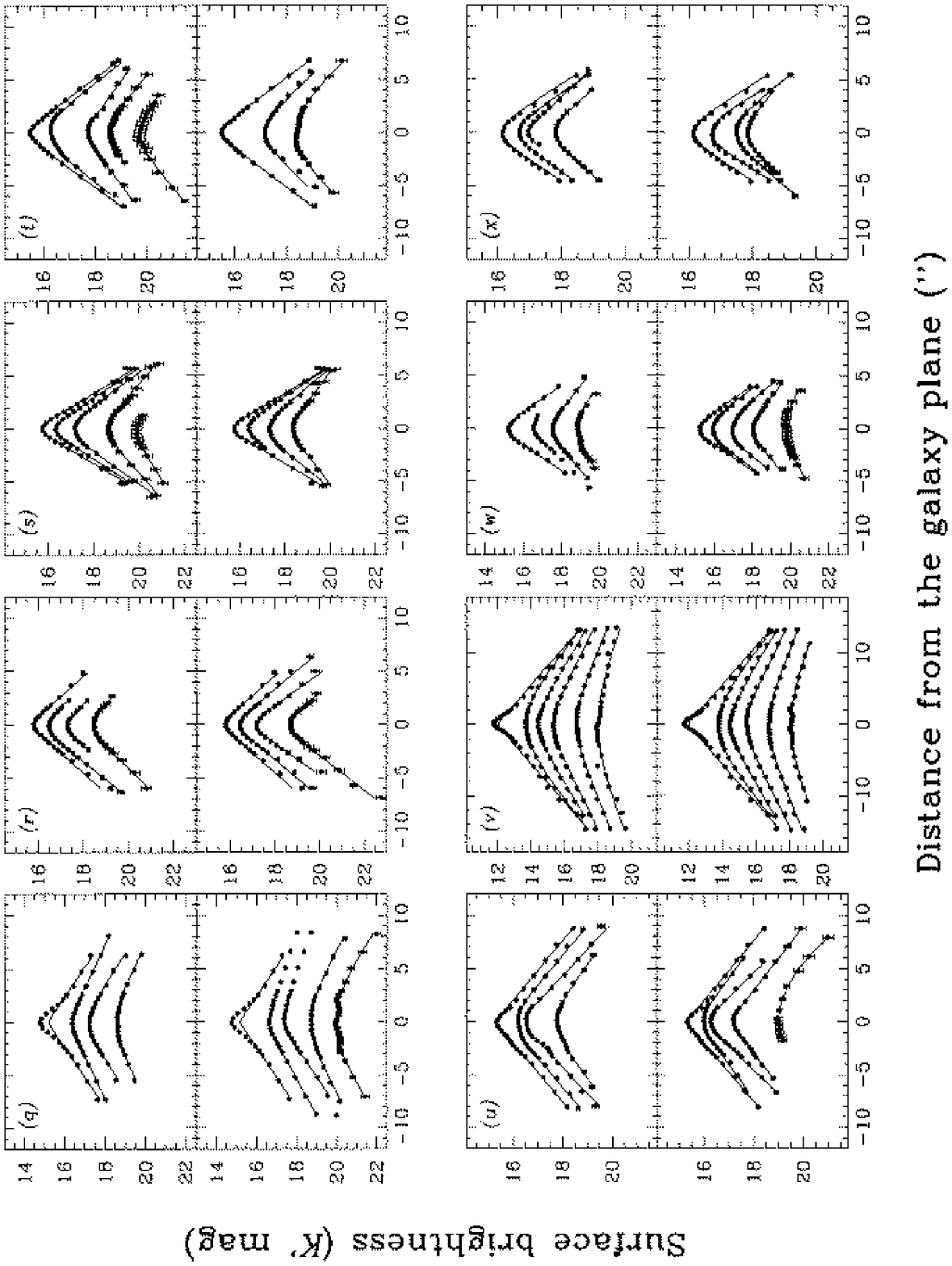}
\caption{}{(Continued) {\it (q)} ESO437G-62; {\it (r)} ESO446G-18; {\it
(s)} ESO446G-44; {\it (t)} ESO460G-31; {\it (u)} ESO487G-02; {\it (v)}
ESO500G-24; \\ {\it (w)} ESO509G-19; {\it (x)} ESO564G-27.} 
\end{figure*}

{
\begin{table}
\caption[ ]{\label{positions.tab}{\bf Positions along the major axis at
which the vertical profiles of Fig. 5
%\ref{fitprofs.fig} 
were extracted.}
\newline Columns: (1) Profile number (on either side of the galaxy
center; 1 = central profile); (2) center position of radially averaged
bin along the major axis (arcseconds); (3) bin width (arcseconds).}

\begin{center}
\begin{tabular}{ccc}
\hline
\hline
Profile number & Center position & Bin width \\
(1) & (2) & (3) \\
\hline 
1 &~~0.0 & 10.8 \\
2 & 11.8 & 12.9 \\
3 & 26.0 & 15.4 \\
4 & 43.1 & 18.7 \\
5 & 63.5 & 22.3 \\
6 & 88.0 & 26.6 \\
\hline
\end{tabular}
\end{center}
\end{table}
}

\begin{figure*}
\hspace*{-1cm}
\psfig{figure=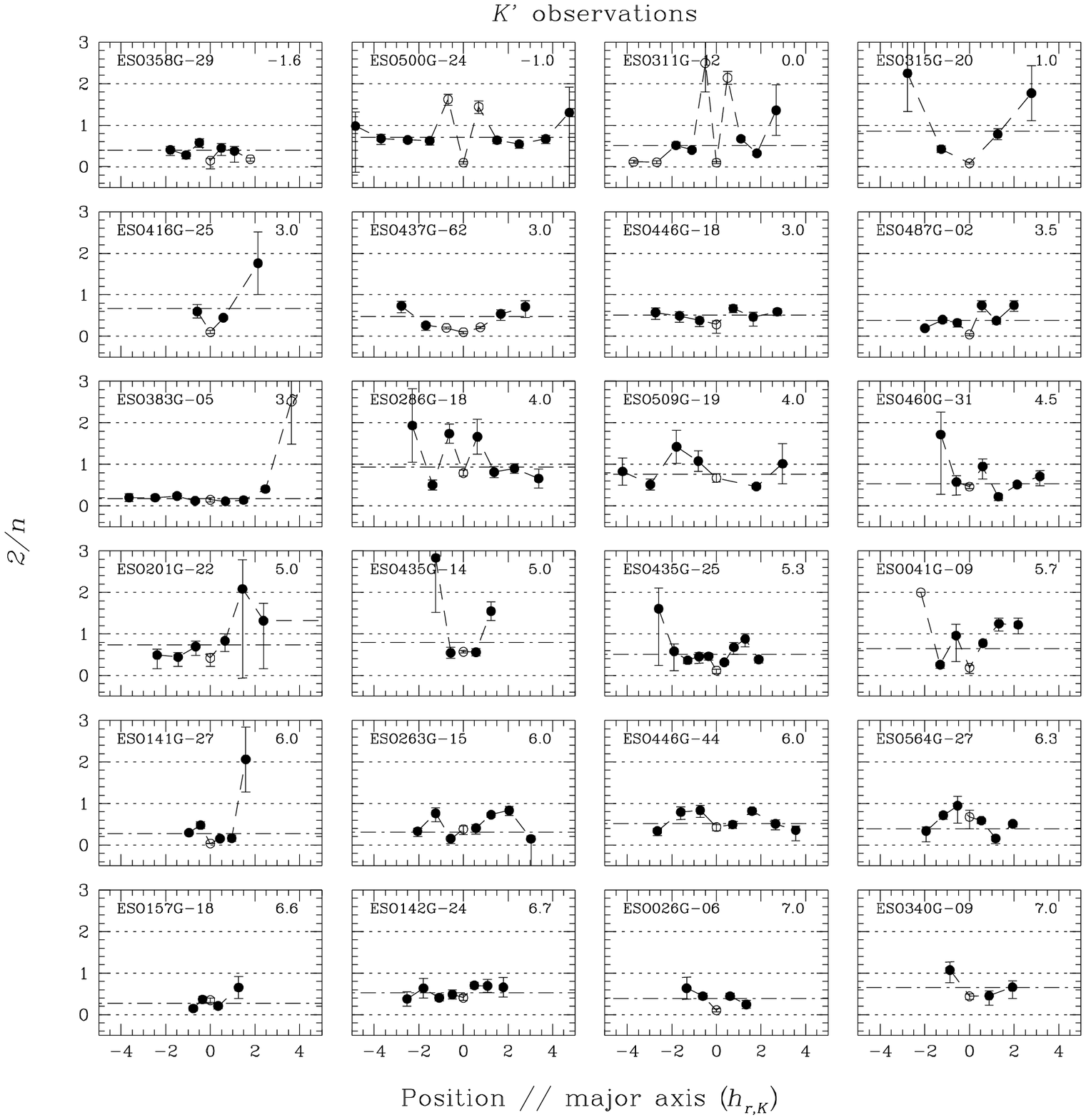}
\caption{}{\label{kband.fig}Best-fitting exponents determining the
sharpness of the peak in the galaxies' vertical luminosity profiles. 
The galaxies are ordered by revised Hubble type, which is shown in the
upper right corner of each panel.  The dotted lines indicate the levels
corresponding to an exponential profile (2/{\it n} = 0), an isothermal
profile (2/{\it n} = 2), and the intermediate sech({\it z}) distribution
(2/{\it n} = 1).  The dash-dotted line represents the mean best-fitting
value for the galaxies' disks, obtained by averaging the filled data
points; the open circles were not taken into account when determining
these levels, because of either bulge light contamination or an
excessively high $\chi^2$ value. The positions along the major axis at 
which the luminosity profiles were extracted are expressed in units of 
their {\it K}-band radial scale length, $h_{r,K}$.} 
\end{figure*}

In Fig.  \ref{kband.fig} we present the distribution of the best-fitting
values for the exponent of the family of luminosity laws (\ref{family.eq})
for our total sample of $K'$-band images. 

We notice that, apart from the rather unpredictable behaviour in the
central areas, where the disk luminosity is contaminated by a
non-negligible bulge light contribution, the vertical luminosity
profiles generally exhibit little variation of 2/{\it n} with position
along the major axis.  As has been shown by Andredakis et al.  (1995),
the early-type bulges are characterized by very steeply peaked
luminosity profiles, whereas in the later-type galaxies the bulge
profiles become closer to exponential.  Therefore, we argue that the
dominant bulge light contribution causes an increase of the cuspiness of
the profiles, especially in the earlier-type galaxies. 

In all panels of Fig.  \ref{kband.fig}, we have indicated an estimated
mean level of the best-fitting 2/{\it n} values by a dash-dotted line. 
These values are given in Table \ref{nvalues.tab}.  
In determining these mean levels, we have excluded both those fits that
were substantially affected by bulge light contamination and those fits
for which the resulting $\chi^2$ values of the minimization routine were
excessively high, due to either a low signal-to-noise ratio or
foreground stars superimposed on the galaxy.  Especially in the
(radially) outer parts of the galaxy disks our $K'$ observations are
rather noisy and therefore unreliable. 

{
\begin{table}

\caption[ ]{\label{nvalues.tab}{\bf Best-fitting 2/{\it n} values}
\newline Columns: (1) Galaxy name (ESO-LV); (2) Revised Hubble Type
({\it T}); (3) and (4) Mean 2/{\it n} values and the (1 $\sigma$) error
obtained from the $K'$-band observations.}

\begin{center}
\begin{tabular}{crcc}
\hline
\hline
Galaxy & Type & $\langle$(2/{\it n})$\rangle_{K'}$ & $\pm$ \\
(1) & (2)~ & (3)  & (4) \\
\hline 
ESO026G-06 &  7.0 & 0.39 & 0.10 \\
ESO041G-09 &  5.7 & 0.64 & 0.17 \\
ESO141G-27 &  6.0 & 0.27 & 0.10 \\
ESO142G-24 &  6.7 & 0.53 & 0.10 \\
ESO157G-18 &  6.6 & 0.27 & 0.09 \\
ESO201G-22 &  5.0 & 0.74 & 0.24 \\
ESO263G-15 &  6.0 & 0.31 & 0.13 \\
ESO286G-18 &  4.0 & 0.93 & 0.30 \\
ESO311G-12 &  0.0 & 0.51 & 0.15 \\
ESO315G-20 &  1.0 & 0.86 & 0.34 \\
ESO340G-09 &  7.0 & 0.65 & 0.16 \\
ESO358G-29 &--1.6 & 0.40 & 0.08 \\
ESO383G-05 &  3.7 & 0.17 & 0.04 \\
ESO416G-25 &  3.0 & 0.67 & 0.21 \\
ESO435G-14 &  5.0 & 0.79 & 0.27 \\
ESO435G-25 &  5.3 & 0.51 & 0.12 \\
ESO437G-62 &  3.0 & 0.48 & 0.16 \\
ESO446G-18 &  3.0 & 0.51 & 0.08 \\
ESO446G-44 &  6.0 & 0.52 & 0.14 \\
ESO460G-31 &  4.5 & 0.53 & 0.21 \\
ESO487G-02 &  3.5 & 0.38 & 0.13 \\
ESO500G-24 &--1.0 & 0.71 & 0.13 \\
ESO509G-19 &  4.0 & 0.76 & 0.22 \\
ESO564G-27 &  6.3 & 0.39 & 0.16 \\
\hline
\end{tabular}
\end{center}
\end{table}
}

In general, we find that the mean levels for the sharpness of the $K'$
band luminosity peaks indicate that the vertical luminosity
distributions are more peaked than expected for the intermediate
sech({\it z}) function, proposed by van der Kruit (1988), though less
peaked than exponential.  

\begin{figure}
\psfig{figure=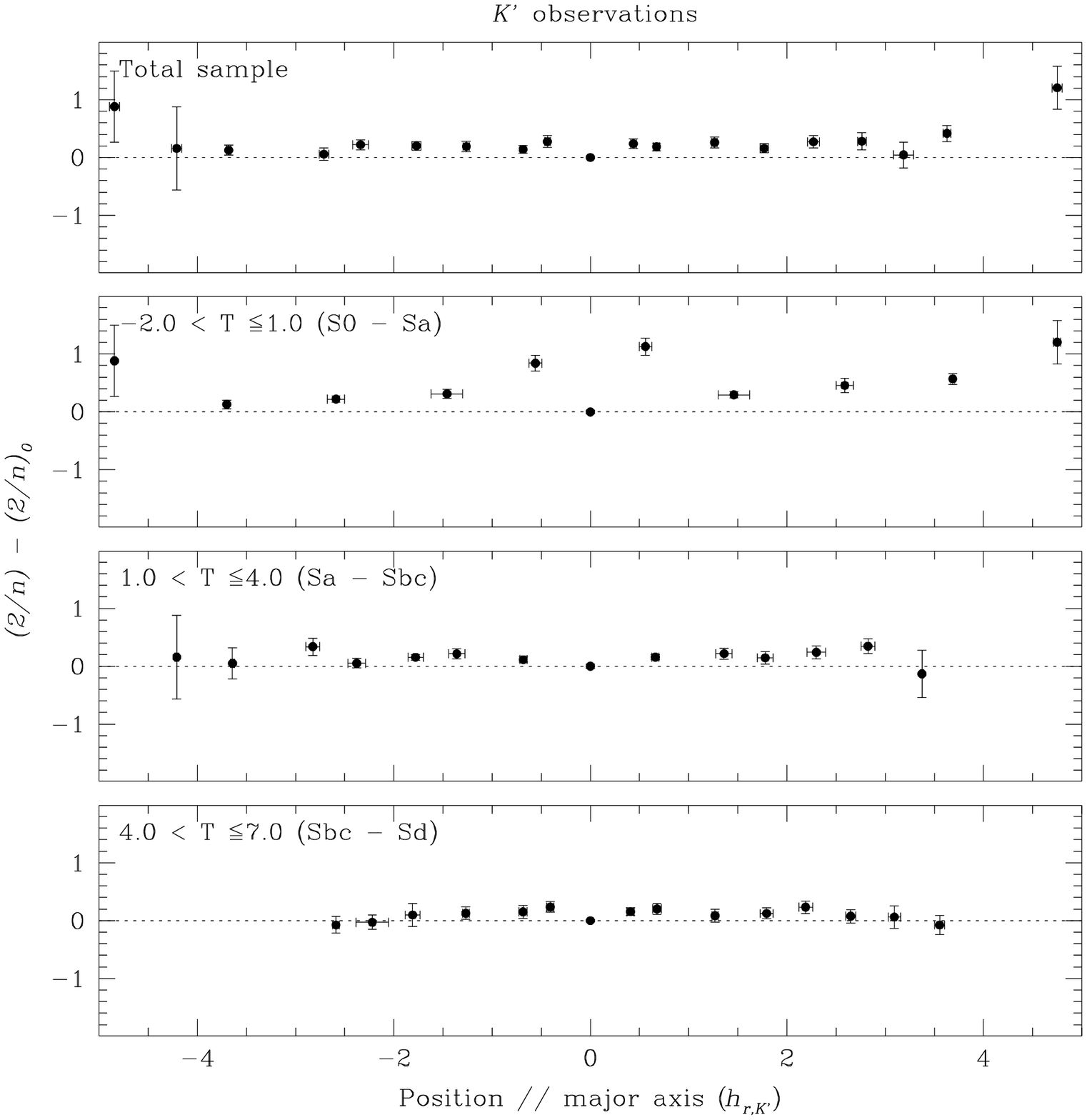,width=9cm}
\caption{}{\label{n-distr.fig}Averaged distributions of the sharpness of
the vertical profiles as a function of position along the major axis,
both for the total $K'$-band sample and for specific bins in revised
Hubble type, {\it T}.}
\end{figure}

To study the variation of the cuspiness of the vertical profiles as a
function of position along the major axis, we have averaged the data
points of our $K'$-band observations in radial bins of 0.5 scale length,
both for our total sample and for subsamples in certain ranges of
revised Hubble type, {\it T}.  The results are shown in Fig. 
\ref{n-distr.fig}.  The central values for 2/{\it n}, $(2/n)_0$, were
used for normalisation; the positions along the major axis are expressed
in units of the {\it K}-band radial scale lengths, $h_{r,K}$.  We
determined the scale lengths by fitting ellipses to the two-dimensional
galaxy isophotes in the regions away from the central dust lane, roughly
between 1 and 4 radial scale lengths, depending on the bulge influence.

We chose to look at the galaxies in the type range $-2.0 < T \le 1.0$ (4
galaxies) because of their clear bulge contribution, and those in the
range $4.0 < T \le 7.0$ (13 galaxies), since they are well-behaved
late-type galaxies without prominent bulges.  The intermediate type
range, with $1.0 < T \le 4.0$, is shown as well.  For these galaxies the
bulge influence is generally small, though not negligible.  In order to
reduce the noise in the earliest-types bin, the radial binning was done
in intervals of 1.0 scale length. 

We notice that the distribution of the sharpness of the vertical
profiles remains, within the errors, constant as a function of position
along the major axis, irrespective of galaxy type.  Some of the
(radially) outermost profiles show a slight increase of 2/{\it n},
corresponding to a rounder vertical profile, but in those regions the
number of useful profiles is small and hence the errors are large.  The
innermost profiles, especially those in the earliest-types bin, are
affected by bulge light and should therefore be left out of the
analysis. 

\subsection{Effects of seeing}

Since we are particularly interested in the behaviour of the vertical
light distributions in edge-on disk galaxies at small {\it z}, we have
to be careful to account for the effects of atmospheric smearing.  No
matter what the exact light distribution is, atmospheric seeing causes
the profiles to show a flat-topped distribution in the galaxy planes. 
In order to circumvent this problem, we have to convolve our models with
the seeing profile.  Assuming that the actual point spread function can
be adequately described by a Gaussian, the seeing convolved model
profile is:

\begin{eqnarray}
K_{\rm c} (z) = \sigma^{-2} \exp(-z^2/2\sigma^2) \int_0^\infty K(x)
I_0(xz/\sigma^2) \nonumber \\
\times \exp(-x^2/2\sigma^2) x \hbox{d}x , \qquad\qquad\qquad\qquad
\end{eqnarray}
where $K_{\rm c}(z)$ and {\it K(z)} are the corrected and intrinsic
model intensity profiles, $\sigma$ is the dispersion of the seeing
Gaussian and $I_0$ is the zeroth-order modified Bessel function of the
first kind (Pritchet \& Kline 1981; Andredakis \& Sanders 1994). 

\begin{figure}
\vspace{-4.4cm}
\psfig{figure=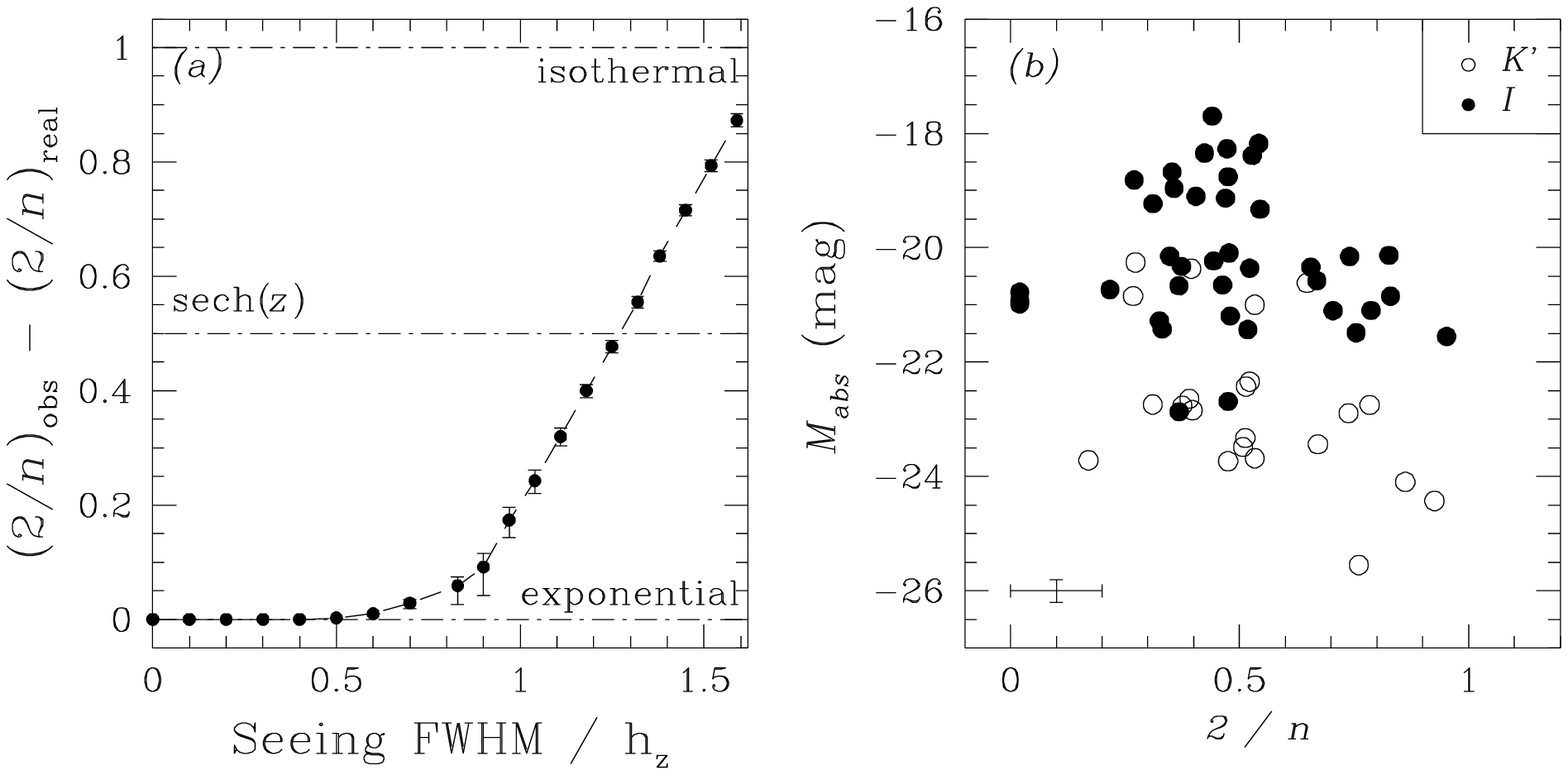,width=9cm}
\caption{}{\label{seeing.fig}{\it (a)} The influence of atmospheric
smearing on an exponential vertical luminosity profile.  The difference
between the true and the observed distributions is shown as a function
of the seeing FWHM, in units of the profile's scale height; {\it (b)}
Distribution of absolute magnitude versus 2/{\it n}, the typical error
size is indicated in the lower left corner.}
\end{figure}

The importance of atmospheric smearing on a particular vertical profile
depends on the ratio of the seeing FWHM and the vertical scale height of
the profile.  The effects are most noticeable for (nearly) exponential
profiles.  In the case of an exponential light distribution, the seeing
convolution becomes non-negligible for those observations with seeing
FWHM $\ge 0.6\; h_z$, see Fig.  \ref{seeing.fig}a.  In the present study
of near-infrared observations, the seeing FWHM varies between 0.12 and
0.57 $h_z$.  Therefore we conclude that the (variation of the) cuspiness
of the observed vertical profiles is real rather than an artifact of the
method applied.  This conclusion is supported by the fact that, in some
of the earliest-type galaxies, we see a variation of the 2/{\it n}
parameter along the major axis due to bulge light contamination, causing
more sharply peaked vertical profiles, which is likely to be real. 

\subsection{Comparison with {\it I-}band observations}

\begin{figure}
\vspace{-3.5cm}
\psfig{figure=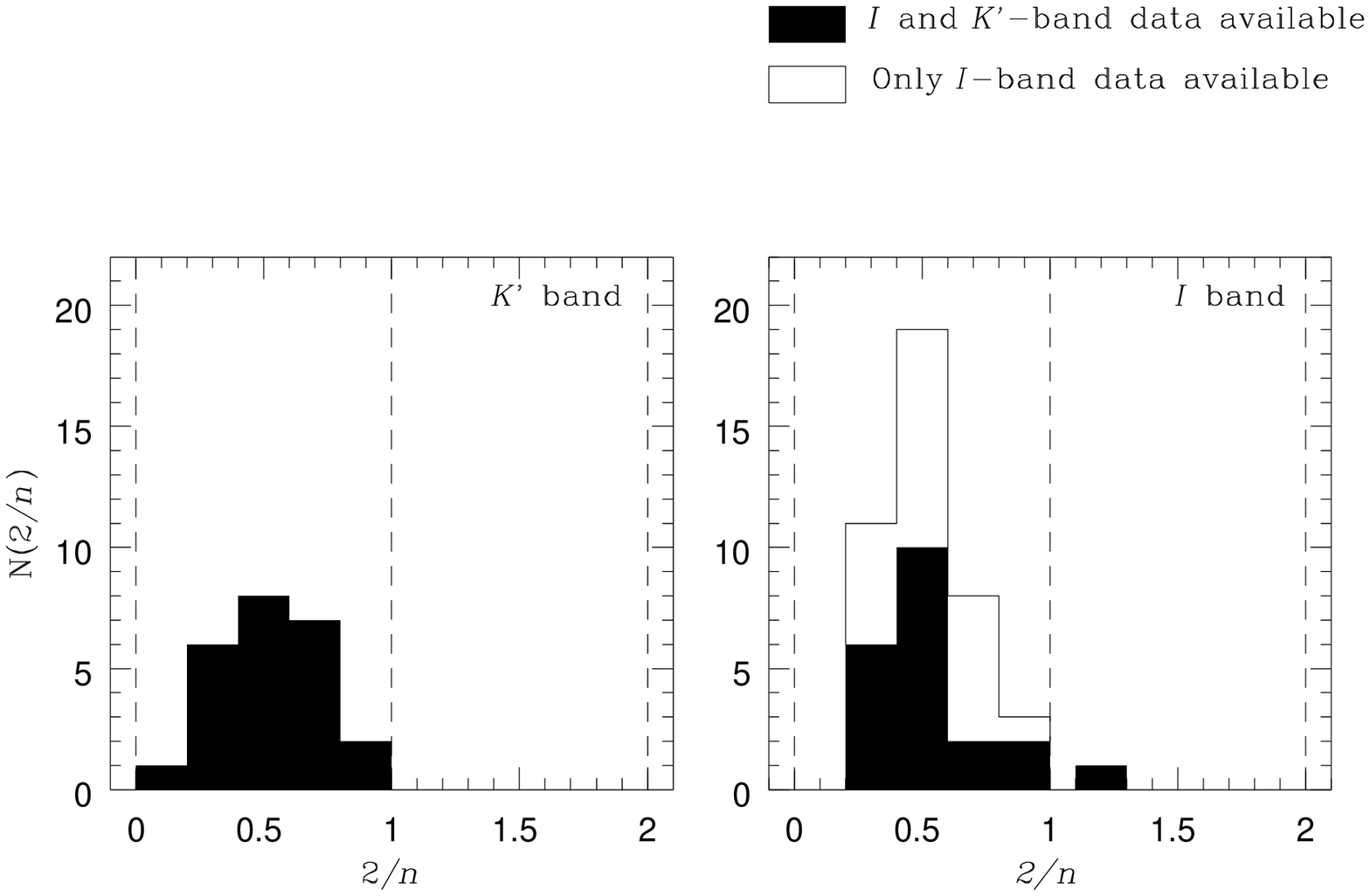,width=9cm}
\caption{}{\label{hist.fig}Comparison between the best-fitting exponents
(2/{\it n}) obtained from both $K'$-band observations (left panel) and
{\it I}-band observations (right panel).  In the right panel, we show
the subsample corresponding to the $K'$ subsample shaded differently. 
Statistically, these two distributions are indistinguishable.}
\end{figure}

To be able to compare our results with those published previously, we
applied this method to our total sample of {\it I}-band observations,
thereby using the same fitting ranges as for the $K'$-band data.  It is
immediately obvious that the {\it I}-band results are much noisier than
those obtained in the $K'$ band, largely due to the presence of clear
dust lanes along the galaxies' major axes.  In those galaxies that show
a well-behaved dependence of the 2/{\it n} value as a function of
position along the galaxy's major axis, we cannot distinguish
statistically between the best-fitting 2/{\it n} values in the {\it I}
band and those obtained in the $K'$ band, within the errors.  This is
made clear in Fig.  \ref{hist.fig}, where we compare the distributions
of best-fitting 2/{\it n} levels between the two passbands.  If we
approximate the distributions of 2/{\it n} in the {\it I} and $K'$ bands
by a Gaussian, we find $\langle$2/{\it n}$\rangle_{K'} = 0.538,
\sigma_{K'} = 0.198$ and $\langle$2/{\it n}$\rangle_I = 0.528, \sigma_I
= 0.202$.  However, $K'$-band observations are clearly preferable to
determine the sharpness of the peak of the vertical luminosity profiles
unambiguously, because of their relative insensitivity to contamination
by galactic dust.  In the presence of a central dust lane, i.e.  a deep
trough in the vertical light distribution, one has to remove the central
regions from the fit to get meaningful answers.  While doing this, one
often ends up only with the exponential outer parts of the vertical
distribution, forcing the fitting routine to yield a more sharply peaked
solution.  Although we tried to fit the vertical profiles over the same
range in {\it z} for both the {\it I} and $K'$-band observations, this
was not always possible due to the disturbing effects of dust in the
{\it I}-band data.  Therefore the best-fitting {\it I}-band 2/{\it n}
values are not very reliable.  In $K'$ the best-fitting 2/{\it n} value
is hardly affected by a different fitting range, since the fit is
dominated by the inner data points, which represent the luminosity
distribution in the galaxy plane. 

From Fig.  \ref{hist.fig} it can be seen that the median of the
best-fitting 2/{\it n} value lies in between the exponential (2/{\it n}
= 0.0) and the intermediate sech({\it z}) model (2/{\it n} = 1.0).  It
is therefore not surprising to note that in previously published papers
by, e.g., Wainscoat et al.  (1989) for the large southern edge-on galaxy
IC 2531, and Gilmore \& Reid (1983) for the Galaxy, the exponential and
the sech({\it z}) models were found to fit the data equally well (van
der Kruit 1988). 

When studying the relation between the sharpness of the peaks and
various global parameters for our sample galaxies, little correlation was
detected, although it seems that the smallest galaxies generally show a
more sharply peaked vertical distribution than the larger galaxies. In
other words, for those galaxies with the faintest absolute magnitudes
($M_I > -20$ or $M_K > -22$), we find a lack of ``rounder'' profiles
($2/n > 0.6$) compared with those galaxies with brighter absolute
magnitudes, as can be seen in Fig. \ref{seeing.fig}b.

Although we do not find clear correlations with global galaxy
parameters, we argue that the radial distribution of 2/{\it n} is
universal (i.e., we do not find any significant variations as a function
of galaxy type), exhibiting only local variations due to, e.g., residual
dust contamination (see, e.g., Wainscoat et al. 1989; Aoki et al. 1991). 

\subsection{Extinction analysis}

In Fig.  \ref{centrprofs.fig} we show the central profiles of all sample
galaxies in both {\it I} and $K'$ band, and their corresponding {\it
I--K} colour, to give an indication of the importance of contamination
by dust in the {\it I}-band observations. 

The vertical colour profiles seem to indicate strong extinction in the
plane of the galaxy, and no extinction in the outer parts.  We therefore
define the constant colour in the outer parts to be the colour of the
stars in the galaxy, and the maximum {\it I--K} difference to be the
{\it I--K} colour excess, $E(I-K)$. 

In Table \ref{extinction.tab} 
we give, on the first line, crude estimates for the maximum (edge-on)
{\it I--K} colour excess, $E(I-K)_{\rm max}$, in the galaxy planes, at
various positions along the galaxies' major axes.  We obtained these
estimates by averaging our vertical colour profiles on both sides of the
galaxy centers.  In deriving these colour excesses we assume that there
is little or no dust mixed in with the stellar population outside the
dust lane region, as seems likely from the smooth vertical colour
profiles.  From a comparison with published colours of moderately
inclined Sc galaxies, Kuchinski \& Terndrup (1996) have shown that there
is indeed little or no reddening away from the dust lane.  Secondly, we
have neglected any change in intrinsic colour due to a stellar
population gradient by assuming that the intrinsic colours of the
dominant stellar population in the dust lane region are identical to
those away from the dust lane.  Kuchinski \& Terndrup (1996) argue that
a population gradient would result in a very small colour change
compared to the total colour change across the dust lane and that we can
therefore assume that the colour exces is largely due to dust
extinction. 

By using the Galactic extinction law we derive that $E(I-K) = 0.77
A_{I,90^\circ}$, the maximum (edge-on) {\it I}-band extinction.  From
this, one can derive the optical depth, and the optial depth   
in {\it V}: $\tau_{V,90^\circ} \; \sim \; 2 \tau_{I,90^\circ}$.

In all our sample galaxies, $E(I-K)_{\rm max}$ remains constant or
decreases as a function of projected galactocentric distance, which
could be a characteristic of an exponentially decreasing dust
contribution.  Jansen et al.  (1994) find, for a small sample of highly
inclined spiral galaxies, that the maximim extinction in the dust lane
($A_{V,{\rm max}}$) decreases rapidly with increasing galactocentric
distance. 

For each position for which we give the maximum {\it I--K} colour
excess, we also tabulate the inferred maximum {\it I}-band optical
depths, $\tau_{I,{\rm max}}$, assuming a uniform mixture of dust and
stars in the galaxy planes, taken from Walterbos \& Kennicutt (1988):

\begin{equation}
A_I = -2.5\log_{10} \Bigl( {1 - \hbox{e}^{-\tau_I} \over \tau_I} \Bigr)
\end{equation}

For the same optical depth, the uniform mixture of dust and stars causes
less {\it I}-band extinction, $A_I$, than the classical foreground
screen model, because part of the extinction lies behind the source.  In
Table \ref{extinction.tab} 
we have also listed the inferred face-on optical depth estimates.  For
face-on galaxies we have assumed the galaxies to be 9 times less opaque
than in the highly-inclined case, which reflects the difference in path
length through the disk at both positions.  Of course, the face-on
optical depth can only be estimated in a rough way; in reality the
optical depth depends on the geometry of the dust and stars (which we
have assumed to be uniformly mixed), the filling factor of the disk
components (in our approximation we have neglected the possibility of a
patchy dust distribution and the presence of spiral arms), and the
major-to-minor axis ratio, which does not need to be 9 for all our
sample galaxies. 

{
\begin{table}
\caption[ ]{\label{extinction.tab}{\bf {\it I}-band extinction estimates in
the galaxy planes}

\newline Columns: (1) Galaxy name; (2)--(4) {\it first line:} Extinction
estimates at the galaxy center, at 1 and at 2 radial ($K'$-band) scale
lengths.  The accuracy of the numbers is indicative of the quality of
the colour profile; {\it second line:} Inferred maximum {\it I}-band
optical depths, $\tau_{I,{\rm max}}$, assuming a uniform mixture of dust
and stars (edge-on and face-on values).}

\begin{center}
\begin{tabular}{cccc}
\hline
\hline
Galaxy & {\it R = 0} & $R = h_{R,K'}$ & $R = 2 h_{R,K'}$ \\
(1) & (2) & (3) & (4) \\
\hline 
ESO 026 G- 06 & 0.50 $\pm$ 0.05  & 0.5  $\pm$ 0.1   & 0.4 $\pm$ 0.1    \\
              & 1.01 $\;\;$ 0.11 & 1.01 $\;\;$ 0.11 & 0.79 $\;\;$ 0.09 \\
ESO 041 G- 09 & 1.15 $\pm$ 0.15  & 1.0  $\pm$ 0.2   & 0.9 $\pm$ 0.2    \\
              & 2.69 $\;\;$ 0.30 & 2.25 $\;\;$ 0.25 & 1.97 $\;\;$ 0.22 \\
ESO 141 G- 27 & 0.0  $\pm$ 0.1   & 0.0  $\pm$ 0.1   & 0.0 $\pm$ 0.1    \\
              & 0.00 $\;\;$ 0.00 & 0.00 $\;\;$ 0.00 & 0.00 $\;\;$ 0.00 \\
ESO 142 G- 24 & 0.65 $\pm$ 0.10  & 0.6  $\pm$ 0.1   & 0.3 $\pm$ 0.1    \\
              & 1.35 $\;\;$ 0.15 & 1.23 $\;\;$ 0.14 & 0.58 $\;\;$ 0.06 \\
ESO 157 G- 18 & 0.30 $\pm$ 0.10  & 0.25 $\pm$ 0.15  & 0.0 $\pm$ 0.1    \\
              & 0.58 $\;\;$ 0.06 & 0.48 $\;\;$ 0.05 & 0.00 $\;\;$ 0.00 \\
ESO 201 G- 22 & 0.9  $\pm$ 0.1   & 0.9  $\pm$ 0.1   & ~~~~~~$\;\;$$^1)$ \\
              & 1.97 $\;\;$ 0.22 & 1.97 $\;\;$ 0.22 & \\
ESO 263 G- 15 & 1.30 $\pm$ 0.10  & 0.85 $\pm$ 0.10  & 0.6 $\pm$ 0.1    \\
              & 3.18 $\;\;$ 0.35 & 1.86 $\;\;$ 0.20 & 1.23 $\;\;$ 0.14 \\
ESO 286 G- 18 & 1.30 $\pm$ 0.10  & 0.9  $\pm$ 0.1   & 0.5 $\pm$ 0.1    \\
              & 3.18 $\;\;$ 0.35 & 1.97 $\;\;$ 0.22 & 1.01 $\;\;$ 0.11 \\
ESO 311 G- 12 & 0.7  $\pm$ 0.1   & 0.4  $\pm$ 0.1   & 0.0 $\pm$ 0.1    \\
              & 1.46 $\;\;$ 0.16 & 0.79 $\;\;$ 0.09 & 0.00 $\;\;$ 0.00 \\
ESO 315 G- 20 & 2.1  $\pm$ 0.1   & 1.4  $\pm$ 0.2   & 0.7 $\pm$ 0.2    \\
              & 6.91 $\;\;$ 0.76 & 3.52 $\;\;$ 0.39 & 1.46 $\;\;$ 0.16 \\
ESO 340 G- 09 & 0.20 $\pm$ 0.05  & 0.25 $\pm$ 0.10  & 0.0 $\pm$ 0.1    \\
              & 0.38 $\;\;$ 0.04 & 0.48 $\;\;$ 0.05 & 0.00 $\;\;$ 0.00 \\
ESO 358 G- 29 & 0.60 $\pm$ 0.05  & 0.0  $\pm$ 0.1   & 0.0 $\pm$ 0.1    \\
              & 1.23 $\;\;$ 0.14 & 0.00 $\;\;$ 0.00 & 0.00 $\;\;$ 0.00 \\
ESO 383 G- 05 & 2.70 $\pm$ 0.05  & 2.3  $\pm$ 0.1   & 1.4 $\pm$ 0.1    \\
              & 12.02 $\;\;$ 1.32& 8.32 $\;\;$ 0.92 & 3.52 $\;\;$ 0.39 \\
ESO 416 G- 25 & 1.6  $\pm$ 0.1   & 0.9  $\pm$ 0.1   & 0.5 $\pm$ 0.1    \\
              & 4.31 $\;\;$ 0.47 & 1.97 $\;\;$ 0.22 & 1.01 $\;\;$ 0.11 \\
ESO 435 G- 14 & 0.8  $\pm$ 0.1   & 0.8  $\pm$ 0.1   & 0.8 $\pm$ 0.1    \\
              & 1.71 $\;\;$ 0.19 & 1.71 $\;\;$ 0.19 & 1.71 $\;\;$ 0.19 \\
ESO 435 G- 25 & 2.3  $\pm$ 0.1   & 1.8  $\pm$ 0.1   & 1.3 $\pm$ 0.1    \\
              & 8.32 $\;\;$ 0.92 & 5.20 $\;\;$ 0.58 & 3.18 $\;\;$ 0.35 \\
ESO 437 G- 62 & 2.00 $\pm$ 0.05  & 1.0  $\pm$ 0.1   & 0.4 $\pm$ 0.1    \\
              & 6.29 $\;\;$ 0.69 & 2.25 $\;\;$ 0.25 & 0.79 $\;\;$ 0.09 \\
ESO 446 G- 18 & 1.2  $\pm$ 0.1   & 1.2  $\pm$ 0.1   & 0.8 $\pm$ 0.1    \\
              & 2.85 $\;\;$ 0.31 & 2.85 $\;\;$ 0.31 & 1.71 $\;\;$ 0.19 \\
ESO 446 G- 44 & 1.50 $\pm$ 0.10  & 1.10 $\pm$ 0.10  & 0.7 $\pm$ 0.1    \\
              & 3.91 $\;\;$ 0.43 & 2.54 $\;\;$ 0.28 & 1.46 $\;\;$ 0.16 \\
ESO 460 G- 31 & 1.0  $\pm$ 0.1   & 0.4  $\pm$ 0.2   & 0.4 $\pm$ 0.1    \\
              & 2.25 $\;\;$ 0.25 & 0.79 $\;\;$ 0.09 & 0.79 $\;\;$ 0.09 \\
ESO 487 G- 02 & 1.50 $\pm$ 0.05  & 0.9  $\pm$ 0.1   & 0.4 $\pm$ 0.1    \\
              & 3.91 $\;\;$ 0.43 & 1.97 $\;\;$ 0.22 & 0.79 $\;\;$ 0.09 \\
ESO 500 G- 24 & 0.0  $\pm$ 0.1   & 0.0  $\pm$ 0.1   & 0.0 $\pm$ 0.1    \\
              & 0.00 $\;\;$ 0.00 & 0.00 $\;\;$ 0.00 & 0.00 $\;\;$ 0.00 \\
ESO 509 G- 19 & 1.00 $\pm$ 0.10  & 0.8  $\pm$ 0.1   & 0.4 $\pm$ 0.1    \\
              & 2.25 $\;\;$ 0.25 & 1.71 $\;\;$ 0.19 & 0.79 $\;\;$ 0.09 \\
ESO 564 G- 27 & 1.20 $\pm$ 0.05  & 0.8  $\pm$ 0.1   & 0.4 $\pm$ 0.1    \\
              & 2.85 $\;\;$ 0.31 & 1.71 $\;\;$ 0.19 & 0.79 $\;\;$ 0.09 \\
\hline
\end{tabular}
\end{center}
\hspace{0.2cm}Note: $^1)$ $\;\;$ {\it I--K} colour profile too noisy.
\end{table}
}

In Fig.  \ref{ch7fig9.fig}{\it a} we present the mean values for the
maximum {\it I}-band edge-on optical depth as a function of Hubble type
and for different positions.  The scatter in this figure is large due to
the small number of data points available, but also because of the
uncertainties in the galaxy classification inherent to edge-on galaxies. 
Therefore, we can only draw qualitative inferences from the behaviour of
the optical depth as a function of galaxy type.  This behaviour suggests
an increasingly important dust contribution from the lenticular and
early spiral galaxies towards later types, although for the latest
galaxy types the dust content seems to diminish relative to the
intermediate types. 

This result is in qualitative agreement with that of Peletier et al. 
(1994), and Peletier \& Balcells (1997) (see also Peletier \& Balcells
1996), who studied the type dependence of the {\it R} and {\it I} vs. 
$K'$ radial scale length ratios of a sample of some 70 disk galaxies,
both face-on and edge-on.  
From Fig.  \ref{ch7fig9.fig}{\it b} it is clear that the radial colour
gradients, indicated by deviations from unity of this ratio, are
smallest for the early-type sample galaxies, whereas in the later types
they vary considerably. 

Peletier et al.  (1994) have shown that scale length ratios due to
stellar population changes are of order 1.1--1.2 in the blue --
near-infrared range; in the {\it I} vs.  $K'$ range this contribution is
likely to be less.  Therefore, the observed scale length ratios largely
represent the galaxies' dust content. 

Although the scale length ratios indicate a more or less constant dust
content for galaxy types later than about $T=3$ (Sb), the data suggests
that the ratios decrease towards later types, as is also found from the
optical depth measurements. 

\begin{figure*}
\psfig{figure=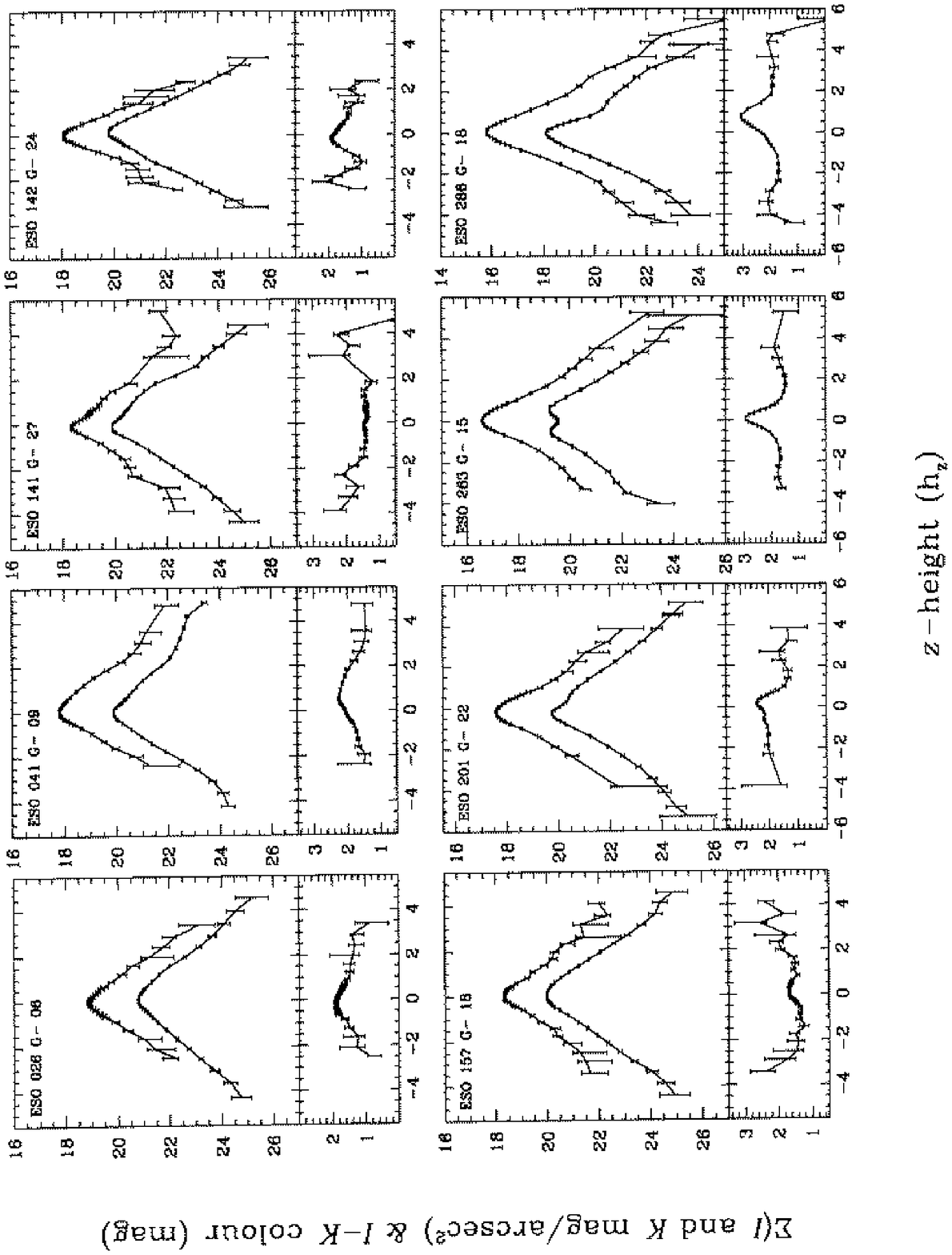}
\caption{}{\label{centrprofs.fig}Central {\it I} and {\it K}-band
vertical profiles for the total sample, and their corresponding {\it I--K}
colours}
\end{figure*}

\begin{figure*}
\addtocounter{figure}{-1}
\psfig{figure=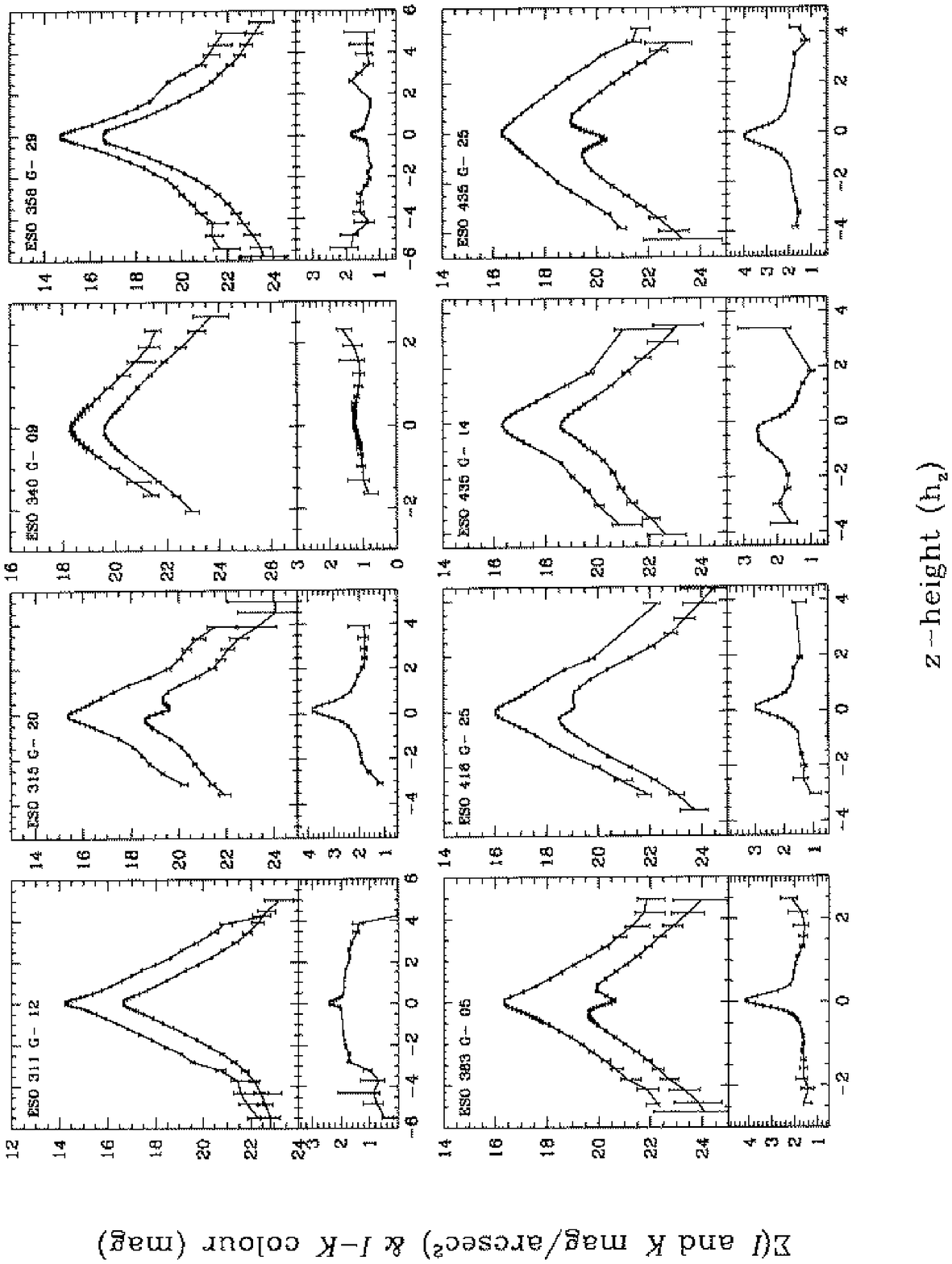}
\caption{}{(Continued)}
\end{figure*}

\begin{figure*}
\addtocounter{figure}{-1}
\psfig{figure=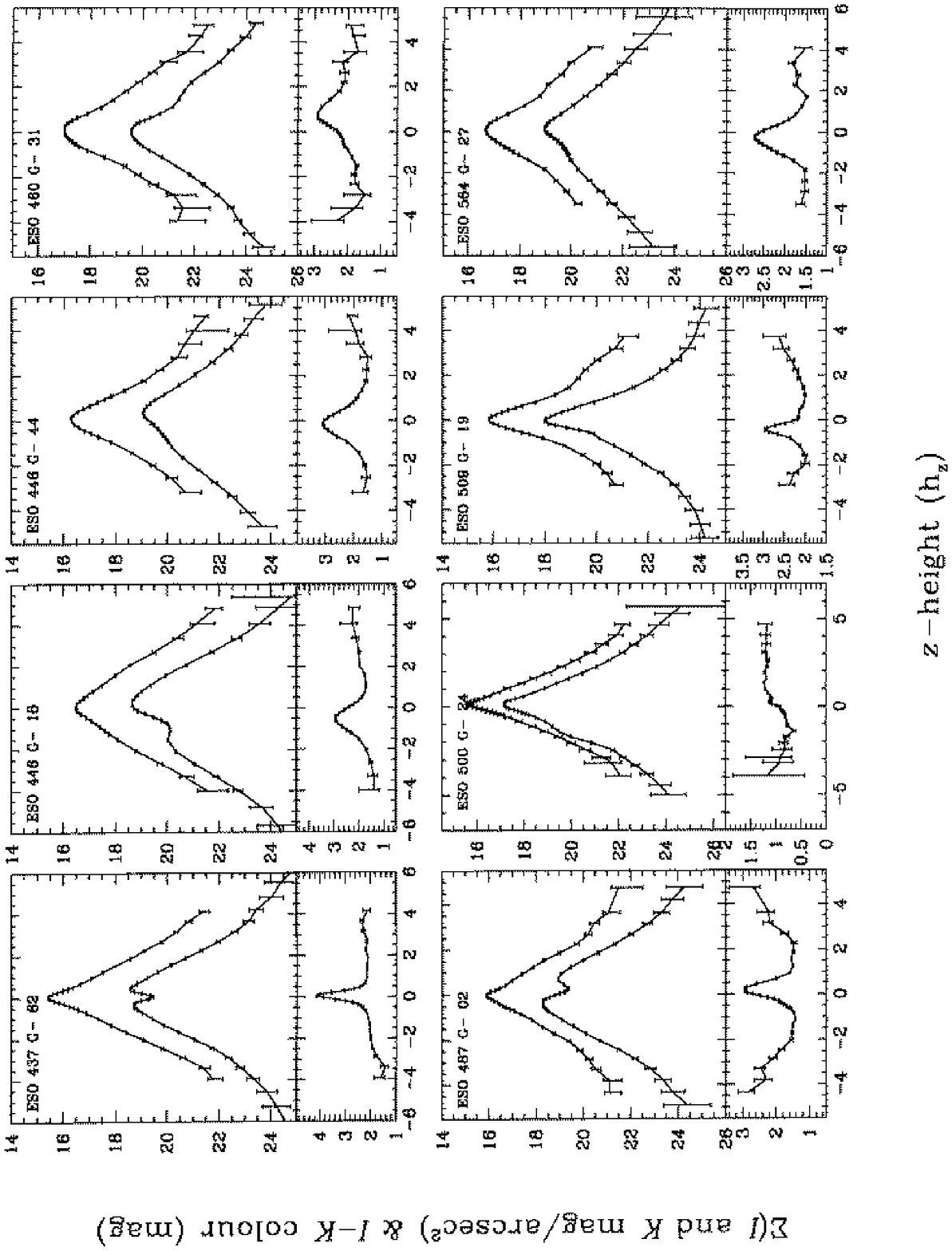}
\caption{}{(Continued)}
\end{figure*}

\begin{figure}
\psfig{figure=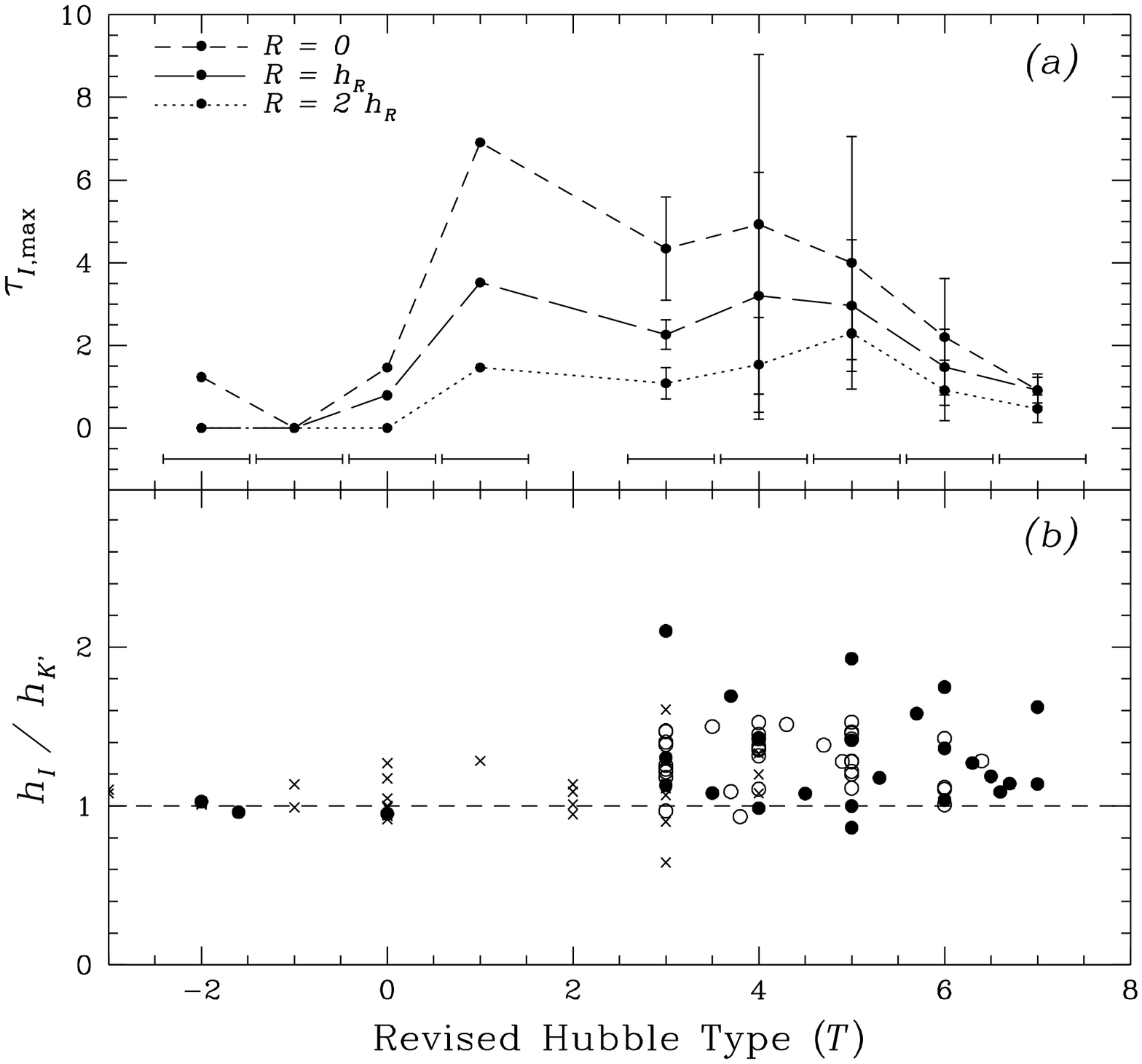,width=8.5cm}
\caption{}{\label{ch7fig9.fig}{\it (a)} Maximum {\it I}-band optical
depths as a function of galaxy type.  The errors are statistical errors;
where no error bars are given, only one data point was available in the
specific type bin, indicated by the bars; {\it (b)} Type dependence
of the {\it I} vs.  $K'$ scale length ratio.  The filled circles are
from the observations presented in this paper, the open circles from
Peletier et al.  (1994), and the crosses from Peletier \& Balcells
(1997).}
\end{figure}

\section{Discussion}
\label{discussion.sec}

\subsection{Main observational results}

As shown in the previous section, our main observational results are
the following:

\begin{itemize}
\item The mean levels for the sharpness of the $K'$-band luminosity
peaks indicate that the vertical luminosity distributions are more
peaked than expected for the intermediate sech({\it z}) distribution,
but rounder than exponential.

\item In the majority of our sample of edge-on disk galaxies, we find
that the sharpness of the peak of the vertical profiles, characterized
by the exponent 2/{\it n} of the generalized family of fitting functions
(\ref{family.eq}) (van der Kruit 1988), varies little with position
along the major axis.  This result is independent of galaxy type. 

\item Due to the unpredictable contamination by in-plane dust, $K'$-band
observations are preferred to {\it I}-band measurements, although
statistically we cannot distinguish between the results obtained in
either of the two passbands.  In general, the {\it I}-band
determinations of the best-fitting 2/{\it n} values are much noisier
than those obtained in the near-infrared. 

\item The behaviour of the maximum {\it I}-band optical depth as a
function of galaxy type suggests an increasingly important dust
contribution from the lenticular and early spiral galaxies towards later
types, although for the latest galaxy types the dust content seems to
diminish relative to the intermediate types. 

\item For those galaxies with the faintest absolute magnitudes, we find
a lack of ``rounder'' profiles compared to those galaxies with
brighter absolute magnitudes.

\end{itemize}

\subsection{Projection effects}

In interpreting the observed surface brightness distributions as surface
density representations, we should be cautious: different stellar
populations may have both different mass-to-light ratios and different
velocity dispersions.  Moreover, the observed galaxy surface brightness
distributions are line-of-sight integrations of the light contributions
and therefore include contributions from very different locations in the
disk.  However, as noted by, e.g., Dove \& Thronson (1993), these
effects are normally considered to be small. 

Kylafis \& Bahcall (1987) showed that the main effects of scattering are
the reduction of the extinction in the dust lane due to forward
scattering and an apparent thickening of the central layer of high
luminosity.  This would not lead to a more sharply peaked profile if the
underlying light (or density) distribution were isothermal.  Moreover,
scattering is not expected to be important in the near-infrared for
edge-on galaxies (Kuchinski \& Terndrup 1996). 

\begin{figure}
\vspace{-4cm}
\psfig{figure=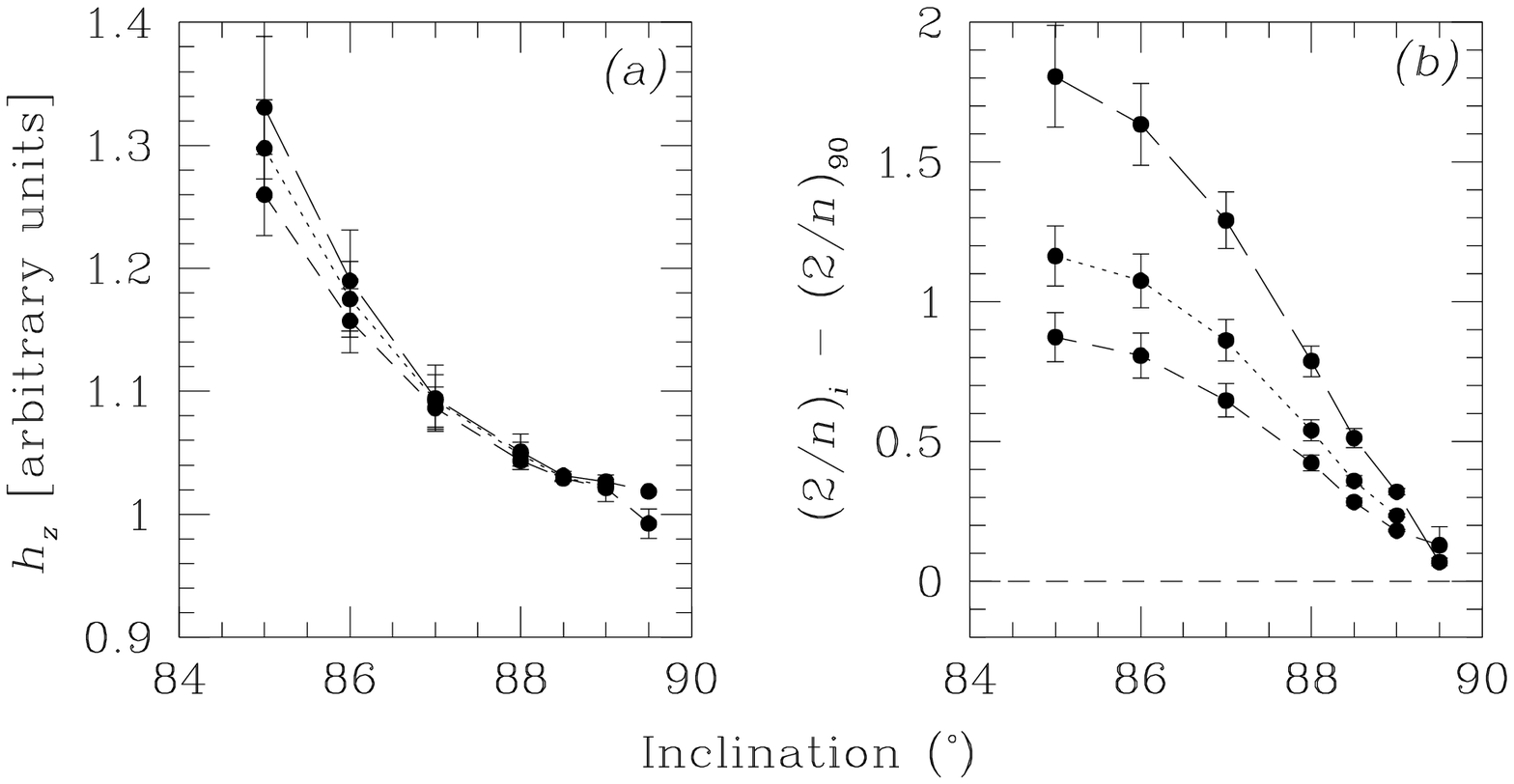,width=9cm}
\caption{}{\label{proj_effects.fig}Importance of projection effects on
the results obtained.  An intrinsic axis ratio of $\bigl( {b \over a}
\bigr)_0 = 0.11$ was assumed (Guthrie, 1992). Projection effects are
shown for a vertical surface brightness profile at the galaxy center
(short-dashed line), at 1 radial scale length (dotted line) and at 2
radial scale lengths (long-dashed line). {\it (a)} Projection
effects cause an increasing scale height with increasing deviations from
a $90^\circ$ inclination; {\it (b)} The influence of projection effects
on the 2/{\it n} parameter.}
\end{figure}

Since the differences between the models discussed here are small, we
have investigated the effects of projection.  Projection effects lead to
an apparent thickening of the high-luminosity midplane of the galaxy. 
Deviations from an inclination of 90$^\circ$ cause a slight increase of
the scale height, as can be seen in Fig.  \ref{proj_effects.fig}a.  The
effect is comparable for the exponential and the isothermal models.  We
are confident that none of our sample galaxies have inclinations lower
than $87^\circ$. 

The effect on the 2/{\it n} values of deviations from a $90^\circ$
inclination is shown in Fig.  \ref{proj_effects.fig}b for an exponential
distribution at the galaxy center, at 1 and at 2 radial scale lengths,
respectively.  For exponential distributions the effects of projection
are most noticeable, compared to more flattened vertical luminosity
distributions.  In fact, the effects of projection can be so large, that
all our galaxies can have intrinsically exponential vertical surface
brightness distributions. 

\subsection{Disk heating}

In this paper we show that the universal vertical structure that is
observed away from the galaxy planes is maintained even when going all
the way down to very low {\it z} distances.  Models that aim to explain
this observational fact should also be able to account for intrinsically
exponential vertical surface brightness or density distributions. 

As Dove \& Thronson (1993) warn, if the vertical distribution of stars
cannot be fit well by the isothermal sheet approximation, which model is
based on physical principles, then rather than invoking an arbitrary
alternative function, a more physical function should be that of a
nonisothermal stellar distribution.  A nonisothermal distribution of
stars can be represented by a linear combination of isothermal
components (Oort 1932).  

It is possible that once a coeval population of stars has been formed,
they do not (or only over a long period of time) interact with other
components and are therefore quasi-independent isothermal components
(Dove \& Thronson 1993).  Such a model would be physically realistic,
since if stars have formed at different times with different velocity
dispersions and have not yet reached an equilibrium state, or if the
stars are dynamically heated during their evolution, the resulting
distribution of stars would not be well approximated by a single
isothermal model.  From observational evidence we know that stellar
subpopulations of different ages have different velocity dispersions. 
Star formation is generally believed to be a continuous process. 
Therefore, a more accurate model for a galaxy disk is a superposition of
a very large number of components, or, as Kuijken (1991) proposed, an
integral representation. 

The observations presented in this paper indicate that the physical
process responsible for the vertical luminosity profiles is probably
both {\em global} and {\em universal} in nature, since we do not find
any significant radial variations of the cuspiness of the profiles, nor
large variations as a function of galaxy type. 

This puts interesting constraints on the dominant vertical heating
mechanism in galaxy disks.  

Although the dominant perturbations in a disk are the spiral waves, they
have a close relationship with the giant molecular clouds (GMCs). 
Julian \& Toomre (1966) and Julian (1967) suggested that the GMCs must
acquire large wakes of material, thus increasing the effective mass of
the combined spiral and GMC perturbation.  These wakes can be very
strong, but will also spread over a large area of the disk, which makes
it more difficult to assess the importance of such wakes in enhancing
the scattering efficiency.  Moreover, at present there is no
satisfactory explanation as to how disk heating, the rate of which must
vary greatly with radius from the observed distribution of GMCs in our
own and other galaxies, can naturally lead to a global and universal
vertical density distribution and a constant scale height with
galactocentric distance (i.e., $z_0$ is independent of {\it R}) (e.g.,
Jenkins 1992). 

\subsection{Isothermal versus exponential distributions}

Jenkins (1992) finds, that his model disk heating process, i.e.,
combined spiral and GMC perturbations enhanced by disk accretion,
together with constant star formation, always leads to a closely
isothermal stellar population.  Wielen (1977) reached a similar
conclusion based on observational data. 

Burkert \& Yoshii (1996) show, based on realistic hydrodynamical
calculations of disk evolution processes, that -- if one starts from a
non-equilibrium gaseous state -- the final vertical stellar density
profile depends strongly on the initial distribution of the protodisk
gas, as opposed to the GMC heating process described in the previous
section. 

On the other hand, if they assume that the gas settles into isothermal
equilibrium prior to star formation and gas cooling, then always an
exponential density profile is formed, although the vertical scale
height increases as a function of decreasing surface brightness.  In
fact, in de Grijs \& Peletier (1997) we presented the results of a
detailed study of the vertical scale height distributions in the present
sample, for which we found an increasing scale height with
galactocentric distance, in particular for the earlier-type galaxies. 
An interesting result from the calculations of Burkert \& Yoshii (1996)
is that when the ratio of the star formation time scale to the cooling
time scale lies in the range between $\sim$ 0.3 and 3, the vertical
stellar density distribution becomes exponential, independent of the
free parameters in their modeling and also independent of the initial
(isothermal) disk temperature and the initial surface density. 

Therefore, the process of crucial importance is that the SFR is adjusted
sooner or later to balance with the local cooling rate (Burkert \&
Yoshii 1996). 

Just et al.  (1996) find that if the SFR decreases with time,
exponential luminosity profiles also grow naturally.  They state that
since the mass-to-light ratio of a stellar population increases with
age, an exponential luminosity profile corresponds to a density profile
that is slightly flattened to the galaxy plane.  However, an isothermal
density distribution is too thick to explain exponential light profiles. 
For a constant SFR an obvious luminosity excess near the plane would
show up in the optical bands, when assuming a heating mechanism of the
type observed in the solar neighbourhood. 

The fact that we observe a more strongly peaked vertical light
distribution than a sech({\it z}) model in all our sample galaxies,
independent of galaxy type, indicates that the process at work here is a
process intrinsic to the disks themselves, rather than a type-dependent
mechanism.  The variations in the cuspiness of our profiles along the
galaxies' major axes are probably due to some local mechanism, e.g.  the
contamination by residual dust.  Although we observe a similar behaviour
in S0s as in later-type galaxies, this does not necessarily mean that
they all possess young populations, although the young population
contributes also in the near-infrared.  The dominant stellar population
in $K'$ is the old population, with ages of several Gyrs, which is
likely present in both early and late-type galaxies.  Therefore our
observational result of a universal vertical density profile in the
near-infrared is not incompatible with no type dependence. 

Finally, the fact that we observe a lack of ``rounder'' profiles in the
smaller galaxies compared with the larger ones, may indicate that we are
hindered by an underlying dust component, which is concentrated towards
the galaxy planes, and more extended in the larger galaxies than in the
smaller ones, at least to an outside observer. It may be that in the
smaller galaxies this dust component affects relatively fewer data points
than in the larger ones, thus causing a bias towards more sharply peaked
vertical profiles.

\section{Summary and Conclusions}
\label{summary.sec}

In this paper we have studied the sharpness of the peaks of the vertical
luminosity and density distribution in a statistically complete sample
of edge-on disk galaxies.  The results obtained in this paper are based
on near-infrared $K'$-band observations.  The main results obtained in
this study are the following:

\begin{itemize} 
\item In the majority of our sample of edge-on disk galaxies, we find
that the sharpness of the peak of the vertical profiles, characterized
by the exponent 2/{\it n} of the generalized family of fitting functions
(\ref{family.eq}) (van der Kruit 1988), varies little with position
along the major axis.  This result is independent of galaxy type. 

\item The mean levels for the sharpness of the $K'$-band luminosity
peaks indicate that the vertical luminosity distributions are more
peaked than expected for the intermediate sech({\it z}) distribution,
but rounder than exponential.  Since projection of a galaxy causes the
profile to be flattened near the galaxy plane, our result is consistent
with the hypothesis that all spiral galaxies have exponential vertical
profiles.  The fact that we observe this in all our sample galaxies
indicates that the process at work here is a process intrinsic to the
disks themselves. 

\item For those galaxies with the faintest absolute magnitudes, we find
a lack of ``rounder'' profiles compared with those galaxies with
brighter absolute magnitudes.  This may indicate that we are hindered by
an underlying dust component, which is concentrated towards the galaxy
planes, and more extended in the larger galaxies than in the smaller
ones, at least to an outside observer.  It may be that in the
smaller galaxies this dust component affects relatively fewer data
points than in the larger ones, thus causing a bias towards more sharply
peaked vertical profiles. 

\end{itemize}

\paragraph{Acknowledgements} - We would like to thank Kor Begeman and
Yiannis Andredakis for their assistance in the development of the
necessary software to carry out this statistical study.  We also
acknowledge the discussions with colleagues at the Kapteyn Institute,
notably Rolf Jansen and David Fisher.

\end{document}